\documentclass[10pt,journal,compsoc]{IEEEtran}

\ifCLASSOPTIONcompsoc
  \usepackage[nocompress]{cite}
\else
  \usepackage{cite}
\fi
\ifCLASSINFOpdf
\else
\fi
\usepackage{amsmath}
\usepackage{array}
\ifCLASSOPTIONcompsoc
 \usepackage[caption=false,font=footnotesize,labelfont=sf,textfont=sf]{subfig}
\else
 \usepackage[caption=false,font=footnotesize]{subfig}
\fi
\usepackage{url}
\newcommand\MYhyperrefoptions{bookmarks=true,bookmarksnumbered=true,
pdfpagemode={UseOutlines},plainpages=false,pdfpagelabels=true,
colorlinks=true,linkcolor={black},citecolor={black},urlcolor={black},
pdftitle={Bare Demo of IEEEtran.cls for Computer Society Journals},%<!CHANGE!
pdfsubject={Typesetting},%<!CHANGE!
pdfauthor={Michael D. Shell},%<!CHANGE!
pdfkeywords={Computer Society, IEEEtran, journal, LaTeX, paper,
             template}}%<^!CHANGE!
\ifCLASSINFOpdf
\usepackage[\MYhyperrefoptions,pdftex]{hyperref}
\else
\usepackage[\MYhyperrefoptions,breaklinks=true,dvips]{hyperref}
\usepackage{breakurl}
\fi
\usepackage{amssymb}
\usepackage{amsthm}
\usepackage{amsfonts}
\usepackage{bm}
\usepackage{booktabs}
\usepackage{multirow}

\usepackage{tikz}
\usepackage{pgfplots}

\pgfplotsset{compat=1.11}
\usetikzlibrary{quotes, angles, shadings, intersections}
\usetikzlibrary{external}
\pgfplotsset{compat=newest}
\usepgfplotslibrary{fillbetween}
\usetikzlibrary{shadows}
\definecolor{red-base}{rgb}{0.925,0.05,0.04}
\definecolor{red-base-dark}{rgb}{0.7,0.2,0.3}
\definecolor{blue-base}{rgb}{0.08,0.52,0.8} % 0.29,0.29,0.91
\definecolor{aqua-base}{rgb}{0.34,0.87,1.0}
\definecolor{green-base}{rgb}{0.1,0.6,0.33}
\definecolor{green-base-shiny}{rgb}{0.17,1.0,0.54}
\definecolor{green-base-dark}{rgb}{0.09,0.5,0.27}
\definecolor{lime-base}{rgb}{0.34,1.0,0.19}
\definecolor{yellow-base}{rgb}{1.0,0.84,0.06}
\definecolor{orange-base}{rgb}{1.0,0.49,0.08}
\definecolor{pink-base}{rgb}{0.98,0.27,1.0}
\definecolor{gray-base}{rgb}{0.88,0.85,0.77}
\definecolor{gray-base-thick}{rgb}{0.78,0.75,0.68}
\definecolor{gray-base-heavy}{rgb}{0.63,0.61,0.55}
\definecolor{gray-base-sombre}{rgb}{0.38,0.36,0.33}
\definecolor{gray-base-dark}{rgb}{0.15,0.25,0.13}

\usepackage[OT2,T1]{fontenc}
\DeclareSymbolFont{cyrletters}{OT2}{wncyr}{m}{n}
\DeclareMathSymbol{\Sha}{\mathalpha}{cyrletters}{"58}

\begin{document}

\title{A Psychoacoustic Quality Criterion for Path-Traced Sound Propagation}

% \author{Chunxiao~Cao,\thanks{State Key Laboratory of CAD\&CG, Zhejiang University, Hangzhou 310058. China.}\\
% {\tt\small ccx4graphics@gmail.com}
% Zili~An,\footnotemark[1]\\
% {\small\url{22021151@zju.edu.cn}}
% Zhong~Ren,\footnotemark[1] \thanks{Corresponding author.}\\
% {\small\url{renzhong@zju.edu.cn}}
% Dinesh~Manocha,\thanks{University of Maryland, College Park, MD 20742, USA.}\\
% {\small\url{dmanocha@umd.edu}}
% Kun~Zhou\footnotemark[1]\\
% {\small\url{kunzhou@zju.edu.cn}}
% }

\author{Chunxiao~Cao,
		Zili~An,
		Zhong~Ren,
		Dinesh~Manocha,
        and~Kun~Zhou% <-this % stops a space
\IEEEcompsocitemizethanks{\IEEEcompsocthanksitem Chunxiao Cao, Zili An, Zhong Ren and Kun Zhou are with the State Key Laboratory of CAD\&CG, Zhejiang University, Hangzhou, China, 310058. \protect\\
Email: ccx4graphics@gmail.com, 22021151@zju.edu.cn,  \protect\\
renzhong@zju.edu.cn, kunzhou@zju.edu.cn.
\IEEEcompsocthanksitem Dinesh Manocha is with the Department of Computer Science, University of Maryland. 8125 Paint Branch Drive, College Park, MD 20742. \protect\\
Email: dmanocha@umd.edu.
}% <-this % stops an unwanted space
% \thanks{Manuscript received April 19, 2005; revised August 26, 2015.}
}

\markboth{IEEE TRANSACTIONS ON VISUALIZATION AND COMPUTER GRAPHICS,~Vol.~XX, No.~X, XXX~XXXX}%
{Shell \MakeLowercase{\textit{et al.}}: Bare Demo of IEEEtran.cls for Computer Society Journals}



\IEEEtitleabstractindextext{%
\begin{abstract}
In developing virtual acoustic environments, it is important to understand the relationship between the computation cost and the perceptual significance of the resultant numerical error. In this paper, we propose a quality criterion that evaluates the error significance of path-tracing-based sound propagation simulators. We present an analytical formula that estimates the error signal power spectrum. With this spectrum estimation, we can use a modified Zwicker's loudness model to calculate the relative loudness of the error signal masked by the ideal output. Our experimental results show that the proposed criterion can explain the human perception of simulation error in a variety of cases.
\end{abstract}

}

\maketitle

\IEEEdisplaynontitleabstractindextext

% \fi
\IEEEpeerreviewmaketitle

\IEEEraisesectionheading{\section{Introduction}\label{sec:introduction}}

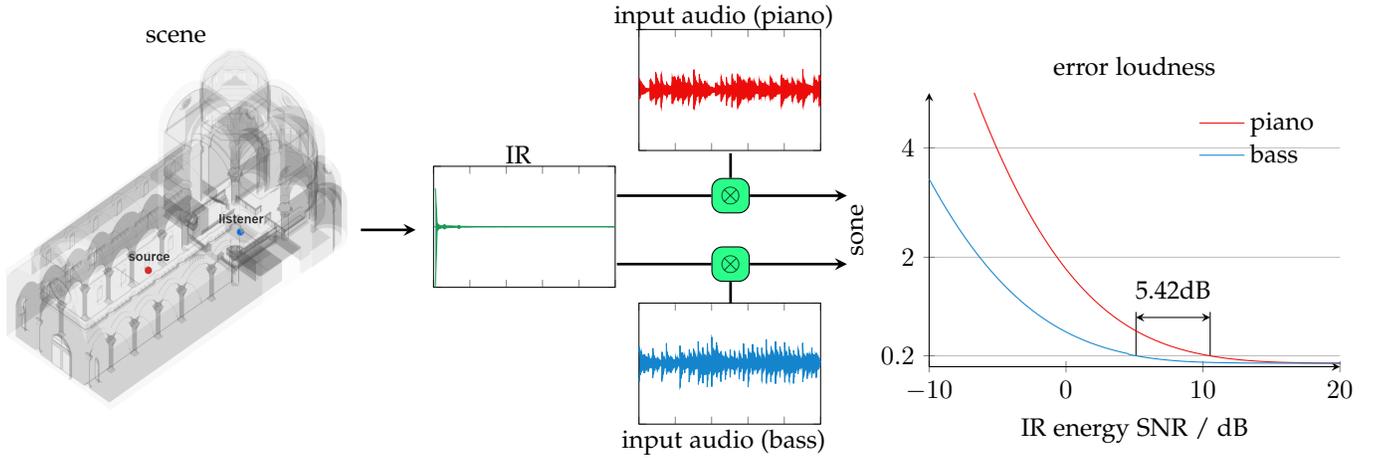
\begin{figure*}[htbp]
	\centering
	\tikzstyle{proc} = [rectangle, rounded corners, minimum width=2, minimum height=2, draw=black, fill=green-base-shiny]
		\begin{tikzpicture}
			\node(IR-scene) at (-0.55\textwidth,0.1\textwidth) {
				\includegraphics[width=0.25\textwidth]{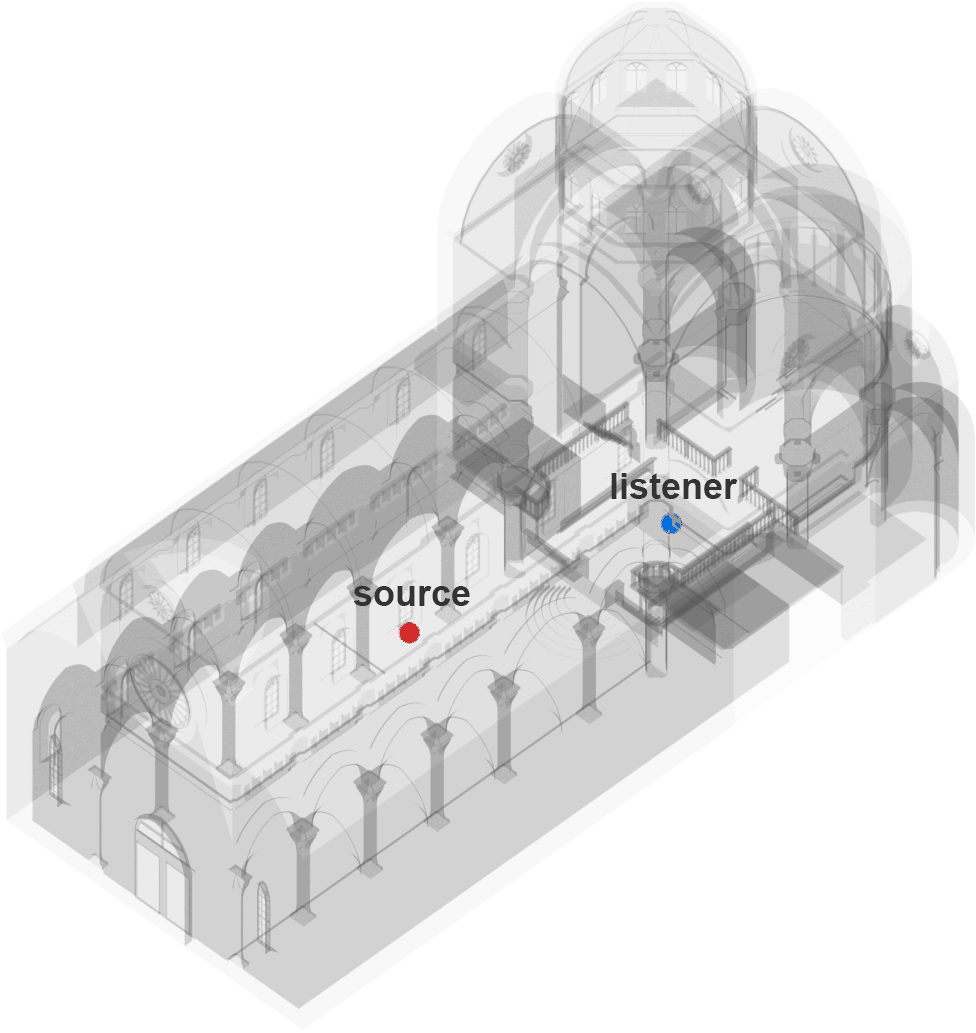}
			};
			\node(IR-scene-label)[above of=IR-scene,yshift=45] {scene};
			\node(IR-wave) at (-0.3\textwidth,0.1\textwidth) {
				\includegraphics[width=0.15\textwidth]{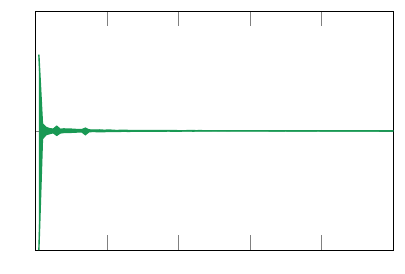}
			};
			\node(IR-wave-label)[above of=IR-wave] {IR};
			\draw[-stealth,line width=1pt] (-0.415\textwidth,0.1\textwidth)--(-0.375\textwidth,0.1\textwidth);
			\node(piano-wave) at (-0.15\textwidth,0.2\textwidth) {
				\includegraphics[width=0.15\textwidth]{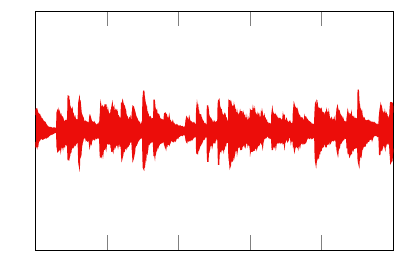}
			};
			\node(piano-wave-label)[above of=piano-wave] {input audio (piano)};
			\node(bass-wave) at (-0.15\textwidth,-0.0\textwidth) {
				\includegraphics[width=0.15\textwidth]{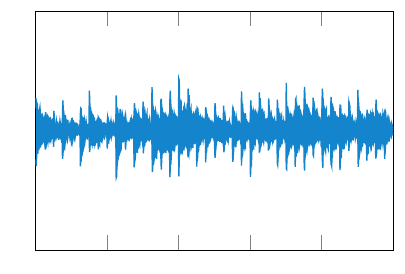}
			};
			\node(bass-wave-label)[below of=bass-wave] {input audio (bass)};
			\node(conv1)[proc] at (-0.145\textwidth,0.125\textwidth){$\otimes$};
			\node(conv2)[proc] at (-0.145\textwidth,0.075\textwidth){$\otimes$};
			\draw[line width=1pt] (-0.145\textwidth,0.156\textwidth)--(conv1);
			\draw[line width=1pt] (-0.145\textwidth,0.046\textwidth)--(conv2);
			\draw[line width=1pt] (-0.228\textwidth,0.125\textwidth)--(conv1);
			\draw[line width=1pt] (-0.228\textwidth,0.075\textwidth)--(conv2);
			\draw[-stealth,line width=1pt] (conv1)--(-0.06\textwidth,0.125\textwidth);
			\draw[-stealth,line width=1pt] (conv2)--(-0.06\textwidth,0.075\textwidth);
	
			\begin{axis}[
				width = 0.3\textwidth,
				height = 0.2\textwidth,
				axis y line = left,
				axis x line = bottom,
				xlabel      = {IR energy SNR / dB},
				ylabel      = {sone},
				ytick = {0.2,2,4},
				xmin = -10, xmax = 20,
				ymin = 0, ymax = 5,
				legend cell align = left,
				legend pos = north east,
				legend transposed = true,
				legend style = {draw=none, fill=none},
				set layers,
				mark layer=axis background,
				/pgf/number format/.cd,
				scale only axis,
				ymajorgrids=true
			]
			\draw [black] (axis cs:5.1075,0.2) -- (axis cs:5.1075,1);
			\draw [black] (axis cs:10.5249,0.2) -- (axis cs:10.5249,1);
			\draw [stealth-stealth,black] (axis cs:10.5249,0.9) -- (axis cs:5.1075,0.9);
			\node(length-label) at (axis cs:7.8162,1.4) {5.42dB};
	
			% thresholds, measured on Chunxiao and Zili
			% \draw [red-base] (axis cs:11.2072846185500,0) -- (axis cs:11.2072846185500,5);
			% \draw [red-base] (axis cs:11.6219197489521,0) -- (axis cs:11.6219197489521,5);
			% \draw [red-base] (axis cs:11.8205440132410,0) -- (axis cs:11.8205440132410,5);
			% \draw [red-base] (axis cs:10.8137454070427,0) -- (axis cs:10.8137454070427,5);
	
			% \draw [blue-base] (axis cs:0.135700755251626,0) -- (axis cs:0.135700755251626,5);
			% \draw [blue-base] (axis cs:-1.55935866149062,0) -- (axis cs:-1.559358661490620,5);
			% \draw [blue-base] (axis cs:8.34915173374034,0) -- (axis cs:8.34915173374034,5);
			% \draw [blue-base] (axis cs:3.92345308028473,0) -- (axis cs:3.92345308028473,5);
			% \draw [blue-base] (axis cs:0.947907658710274,0) -- (axis cs:0.947907658710274,5);

			\addplot[red-base] table [x=SNR,y=piano, unbounded coords=discard, col sep=comma] {img/loudness_SNR_relation.csv};
			\addlegendentry{piano};
			\addplot[blue-base] table [x=SNR,y=bass, unbounded coords=discard, col sep=comma] {img/loudness_SNR_relation.csv};
			\addlegendentry{bass};
			\end{axis}
			\node(bass-wave-label) at (0.15\textwidth,0.22\textwidth) {error loudness};
		\end{tikzpicture}
		\caption{Even in the same acoustic environment, human sensitivity to simulation errors in sound propagation may vary for different input audio clips. Our criterion exposes this difference and allows the developer to adjust the simulation quality for different cases. The figure above shows that, to achieve the same error loudness level (0.2 sone), the required IR SNR is 5.42dB higher for the ``piano'' sound when compared to the ``bass'' sound.}
	\label{fig:teaser}
\end{figure*}
  
\IEEEPARstart{I}{n} virtual reality, computational room acoustics, and various other fields, we need to calculate the sound received by human ears in a virtual acoustic scene. High quality sound rendering can significantly improve the overall user experience in virtual reality \cite{Larsson2002BetterPA}. To calculate the received sound, we need to simulate various physical phenomena in the sound generation and propagation process \cite{Liu2020SoundSP}. Simulation of sound generation provides the user information about sound sources \cite{moss2010soundingliquids,Cirio2016CrumplingSS,Li2015InteractiveAT}, and simulation of sound propagation provides critical information about the sound environment \cite{Rungta2016PsychoacousticCO,Blauert1986AuditorySS,Zahorik2005AuditoryDP}.

Sound propagation from the source to the listener can be modeled as a linear time-invariant system, which can be fully described with its impulse response (IR) function $h(t)$, a function calculated by the simulator. Given the input audio $f_i(t)$, the propagated audio $f_o$ received at the listener is given by the equation $f_o(t) = f_i(t) * h(t)$, where the asterisk stands for convolution. All simulation algorithms inevitably generate an error $h_e$ over $h$, adding an extra signal $f_e = f_i * h_e$ to the output audio. The simulation quality can be measured with the energy signal-to-noise ratio (SNR) of $h$.

While some methods simulate sound propagation by solving the wave equation directly \cite{mehra2015wave}. Speed-critical applications usually rely on methods based on geometrical acoustic (GA) methods \cite{lauterbach2007interactive, chandak2008ad-frustum} or hybrid methods \cite{yeh2013waveray}, Among GA methods, Monte Carlo path tracing \cite{cao2016BDPT, schissler2014high-order, schissler2016multisource} is one of the most popular methods for real-time simulation. In path tracing, $h(t)$ is given as an integral calculated by the Monte Carlo method, which uses a large number of random samples called ``paths'' to simulate the propagation process. An advantage of path tracing is that its computation budget can be easily adjusted to accommodate for different quality requirements. For example, there exists a simple relationship between the number of samples and IR SNR, allowing us to balance between speed and quality easily by controlling the sample count. Other quality control methods have also been proposed in previous works \cite{schissler2016multisource, cao2016BDPT}.

Since the output audio is usually received by humans, we would like to know the effect of simulation error on receivers' hearing experiences. Simply put, we need to predict the adequate IR SNR value, under which we can produce plausible outputs for users. However, it turns out that building the relationship between the user experience and IR SNR is not simple. For real-time simulators, where $h$ is calculated on-the-fly and $h_e$ changes in every frame, the problem becomes even more complicated. Unfortunately, the change of $h_e$ between frames is critical for the user to pick up the error of the simulator. There is no analysis on the noise of real-time simulators, let alone its influence on users' hearing experiences.

In this paper, we will show that it is possible to bridge the gap between IR quality and user hearing experience using established psychoacoustic models. The main results in this paper include:
\begin{itemize}
	\item A spectral analysis for the error signal $f_e$ of both static and real-time path-tracing-based simulators. An analytical expression for the power spectrum of $f_e$ is given, along with practical approximations. With this expression, we can estimate $f_e$ with IR SNR and the spectrum of the input audio.
	\item A criterion that predicts the audibility of the error signal $f_e$ based on the widely-used Zwicker's loudness model \cite{ZwickerLoudness1,ZwickerLoudness2,ZwickerLoudness3}. From Zwicker's original algorithm, which calculates the ``absolute loudness'' of a signal in a quiet environment, we develop an algorithm to calculate the ``relative loudness'' of $f_e$ in the presence of $f_o$. This algorithm allows us to predict the ``audibility thresholds'' below which the error $f_e$ becomes inaudible, which is valuable for render quality control. Our experiments show the validity of our criterion in various cases.
\end{itemize}

Combining the spectral analysis with the loudness model, we achieved a novel method that explains the error perceptibility under various acoustic scenarios with different input audio clips. While the spectral analysis result is dedicated to path-tracing-based simulators, the criterion can be applied to methods other than path tracing if an estimation for their power spectra exists.

\section{Background and Related Work}

To facilitate discussion, we introduce the necessary background knowledge in this section for readers unfamiliar with acoustics and sound simulation. Readers can refer to \autoref{table:symbols} for the definitions of mathematical symbols in this paper.

\subsection{Numeric Characteristics of Audio Signals}

\begin{table}
	\caption{Symbols in this paper. We use square brackets for parameters of random variables, round brackets for functions, and calligraphic letters for operators.}
	\label{table:symbols}
	\begin{center}
		\linespread{3}
		\begin{tabular}{m{0.18\linewidth}m{0.72\linewidth}}
			\toprule
			symbol & explanation \\
         	\midrule
			$E[X]$ & mathematical expectation of variable $X$ \\
			$\sigma[X]$ & standard variance of $X$. $\sigma^2[X]=(\sigma[X])^2$ \\
			$\text{Cov}[X,Y]$ & covariance of $X$ and $Y$ \\
			$\mathbb{R}$ & set of real numbers \\
			$\mathbb{Z}$ & set of integers \\
			$\bar{x}$ & complex conjugate of $x$ \\
			$\delta$ & Dirac delta function \\
			& $\int_\mathbb{R}f(t)\delta(t)dt=f(0)$ \\
			$\mathcal{F}(f)$ & Fourier transform \\
			& $\mathcal{F}(f)(\omega)=\int_\mathbb{R}f(t)e^{-2\pi i\omega t}dt$ \\
			\bottomrule
		\end{tabular}
	\end{center}
\end{table}

\subsubsection{Power and Power Spectrum}

In signal processing, the power of a signal $f$ of duration $T$ is given by the equation $p=\frac{1}{T}\int_\mathbb{R}f(t)^2dt$. For a signal of infinite duration, the power is given as $\lim_{T\to+\infty}\frac{1}{T}\int_{-T/2}^{T/2}f(t)^2dt$. While power is the simplest metric for signal intensity, we prefer to know the power distribution of a signal on its frequency domain, as the human hearing system responds to frequential characteristics of incoming sounds. Such a distribution is given by the power spectrum. For a signal $f$, the power spectrum is given by the equation $\frac{1}{T}|\mathcal{F}(f)|^2$, where $T$ is the duration of the signal.

We can also describe the signal intensity with pressure or power levels. For a signal of power $p$, the power level $L_{p/p_r}$ is given by the equation
\begin{equation}
	L_{p/p_r} = 10\log_{10}\frac{p}{p_r}\,\text{dB}.
\end{equation}
Here $p_r$ is an appointed reference value. For example, decibels relative to full scale (dB FS) is the case where $p_r=1$. Sound pressure level (dB SPL) is the case where the average sound pressure is $2\times 10^{-5}\text{Pa}$ at $p_r$.

For the error signal $f_e$ overlapped on the signal $f$, we assume that $E[f_e]=0$ and that its Fourier spectrum $\mathcal{F}(E[f_e])$ is also zero. Thus we have
\begin{equation}
	\begin{aligned}
	E[\frac{1}{T}|\mathcal{F}(f_e)|^2]&=\frac{1}{T}E[|\mathcal{F}(f_e)-\mathcal{F}(E[f_e])|^2] \\
	&=\frac{1}{T}E[|\mathcal{F}(f_e)-E[\mathcal{F}(f_e)]|^2] \\
	&=\frac{1}{T}\sigma^2[\mathcal{F}(f_e)].
	\end{aligned}
\end{equation}
Thus, the power spectrum of the error is also the spectrum of its variance. This $\sigma^2[\mathcal{F}(f_e)]$ is what we are trying to calculate in \autoref{section:SpectrumAnalysis}.

\subsubsection{Energy SNR}

A useful numerical metric for signal error is the energy signal-to-noise ratio (SNR). For a signal $f$ of limited duration overlapped with an error signal $f_e$, the energy SNR is
\begin{equation}
	\text{SNR}(f)=\frac{\int_{\mathbb{R}}f(t)^2dt}{\int_{\mathbb{R}}f_e(t)^2dt}.
\end{equation}
When the $f_e$ is a random function,
\begin{equation}
	\text{SNR}(f)=\frac{\int_{\mathbb{R}}f(t)^2dt}{E[\int_{\mathbb{R}}f_e(t)^2dt]}.
	\label{eq:SNRdefinition}
\end{equation}
Energy SNR is also frequently given in the form of SNR levels:
\begin{equation}
	L_{\text{SNR}}(f)=10\log_{10}\text{SNR}(f)\,\text{dB}.
\end{equation}

\subsection{Mechanisms Influencing Auditory Perceptibility}

A good way to understand the various factors that influence human sound perception is to classify them by their effects on resultant neuron signals. In this way, we can divide hearing mechanisms into excitation and suppression mechanisms.

The most remarkable feature of the excitation mechanism is frequency selectivity: sound components of different frequencies are separated by the auditory system and perceived by different receptors. Frequency selectivity has been well-studied from both anatomical and psychological perspectives. It has been observed that components of different frequencies resonate with different positions on the basilar membrane (BM) of the cochlea. In psychoacoustics, a rough relationship between frequency and BM position is given by frequency scales like the Bark scale \cite{criticalband1}. On further investigation, one finds that a single point on the BM responds to a range of frequencies rather than a single frequency. We use band-pass filters called auditory filters to describe these frequency response patterns. There are several models describing the characteristics of these filters, with parameters like critical bandwidth (CB) \cite{criticalband1,criticalband2} or equivalent rectangular bandwidth (ERB) \cite{erb1}. With an appropriate scheme, one can divide the audible frequency range into several bands for which the excitations minimally interfere.

The suppression mechanism of human hearing is not as well understood as excitation. An important related discovery is otoacoustic emission \cite{otoacoustic1}, which shows that the cochlea adjusts its sensibility not only neurally, but also mechanically. The suppression mechanism affects frequential sensitivity in a highly nonlinear and nonlocal way and drives many effects such as the ``upward spread'' of masking \cite{upwardspread1}, where a low-frequency sound can suppress the sensibility of another sound with much higher frequency, and auditory adaptation \cite{adaptation1}, where the continued presence of a sound will be regarded as ``background'' and ignored by the hearing system. Other studies show that the frequency selectivity of hearing is also a combined effect of activation and suppression mechanisms \cite{freqselectivity2}.

\subsection{Masking Effect}

The masking effect describes the change of perceptibility of a sound (signal) in the presence of another sound (masker). Due to its relationship with various hearing mechanisms, masking has long been a useful tool in the study of human hearing. The perceptibility of the signal is non-linearly related to its intensity level: it decreases much faster below a certain threshold than above it \cite{masking1}. This threshold is called the mask threshold. Naturally, this threshold should be above the absolute hearing threshold, under which a sound is imperceptible. The absolute threshold can also be considered a special case of mask threshold, with the masker being the internal noise generated by physiological activities \cite{psychoacousticsbook}. Here, we note the absolute threshold as $p_0$ and the mask threshold as $\text{max}\{p_0,p_t\}$, as $p_t$ can usually be described with simple models.

Narrow-band masking occurs when a narrow-band masker masks a signal (usually a sinusoidal one) within the band. This phenomenon is closely related to frequency selectivity. The relationship between $p_t$ of narrow-band masking and the power spectrum of the masker is generally assumed to be linear \cite{masking2}:
\begin{equation}
	p_t = \int_{\mathbb{R}} p_m\cdot H_a d\omega
\end{equation}
where $p_m$ is the power spectrum of the masker and $H_a$ is a function that shows the shape of the auditory filter.

Broad-band masking occurs when the signal is masked by a broad-band masker. When the signal is sinusoidal and the masker is white noise, \cite{psychoacousticsbook} approximates the mask threshold with the relationship $p_t(\omega_t)=k(\omega_t)\cdot p_m(\omega_t)$, where $\omega_t$ is the signal frequency and $k$ is a parameter related only to $\omega_t$. With this relationship, a uniform masking noise is developed in \cite{psychoacousticsbook} that makes $p_t$ equal for all frequencies.

\subsubsection{Loudness}

The perceived intensity of the signal under a certain masker is called its loudness. The ``loudness'' in the general sense is the loudness when the masker is the internal noise. Loudness models are closely related to auditory filter models. Different auditory filter descriptions lead to different loudness algorithms. The criterion in this paper is related to the loudness algorithm proposed by Zwicker, which uses CB to describe auditory filters.

The general unit of loudness is sone. Like power and SNR, there is a corresponding unit for loudness level named phon. Given an audio signal, different loudness algorithms may result in different sone values. Since our criterion is a metric for relative loudness, we also use sone as the unit of its result values. 

\subsection{Sound Propagation and IR}

The goal of sound propagation simulation is to calculate the IR of the sound environment. As its name indicates, IR represents the sound received at the listener's position when the sound source emits an impulse signal.

\subsubsection{The Structure of IR}

The impulse response can be broken down into three parts: direct contribution, early reflections and late reflections. The direct contribution represents the sound propagated from the source to the listener directly. It may not exist when there is an occluding object between the listener and the source. This part usually has an analytical expression and is computed directly in most path-tracing-based simulators. The early reflections represent the sound reflected by the scene in the first few times. It is related to the geometry and acoustic characteristics of objects near the listener. The late reflections represent the sound reflected multiple times in the environment. It is related to the general volume and sound absorption rate of the scene. Late reflections are important to user experience because they give the listener a ``sense of volume'' of the scene \cite{Blauert1986AuditorySS}. In path-tracing-based simulators, both early and late reflections are generated by random sampling and are thus not immune to numerical error.

\subsubsection{Psychoacoustics in IR Computation}

While path tracing is already one of the most efficient methods to calculate IR, the calculation process is still costly, especially for real-time applications and low-end devices. It is possible to use psychoacoustic knowledge to improve the simulation efficiency. For example, Schuitman and Vries use a psychoacoustic method to suppress IR noises \cite{PANoiseReduction}. Rungta et al. reduce the computation cost using the relationship between the scene changes and the audibility of IR changes \cite{PReverb}. Zhou et al. use the A-weighting scheme to estimate the error significance and the appropriate computational budget for simulation \cite{AWeighting}. However, the error analyses in these works are usually based on oversimplified models that ignore some critical factors, like the masking effect and the frame-to-frame IR variance. In contrast, these factors are all taken into consideration in our new criterion.

\section{Spectrum Analysis of the Error Signal}
\label{section:SpectrumAnalysis}

The spectrum of the ideal output $f_o = f_i * h$ is given by the equation $\mathcal{F}(f_o) = \mathcal{F}(f_i)\cdot \mathcal{F}(h)$. In this section, we will estimate the power spectrum of the error signal $f_e$. The calculation process is similar to the proof of the Wiener-Khinchin theorem \cite{wiener}. When dealing with discrete signals that are sampled from continuous signals, we use the equation
\begin{equation}
	\mathcal{F}(f)(\omega)=\sum_{t\in\mathbb{Z}}\frac{1}{s}f(st)e^{-2\pi i\omega st}
\end{equation}
for Fourier transform of discrete functions. Here $s$ is the sample rate. This equation still satisfies $\mathcal{F}(f\cdot g)=\mathcal{F}(f)*\mathcal{F}(g)$ and $\mathcal{F}(f*g)=\mathcal{F}(f)\cdot\mathcal{F}(g)$. Under this definition, we have the discrete version of Parseval's theorem \cite{combfunction}:
\begin{equation}
	\sum_{t\in\mathbb{Z}}\frac{1}{s}f(st)^2dt = \int_{-s/2}^{s/2} |\mathcal{F}(f)(\omega)|^2d\omega.
	\label{eq:DiscreteParseval}
\end{equation}

\subsection{Static Case}
\label{section:static-case}

In path tracing, the IR function is estimated by accumulating a series of independent random samples in the form $a\delta(t-t_0)$. The accumulated result $h'$ can be regarded as a stochastic process with expectation $E[h'] = h$. In practice, $h'$ is a discretely sampled function, and $E[h'(t)]=\int_{t-1/2s}^{t+1/2s}h(t)dt$. When the number of samples is sufficiently large, $h'(t_0)$ and $h'(t_1)$ become approximately independent when $t_0\neq t_1$. This independence is critical to our analysis below.

The sample error $h_e = h'-E[h']$ is also a stochastic process. From the independence assumption, we have $E[h_e]=0$ and $\forall t_0\neq t_1,\text{Cov}[h_e(t_0),h_e(t_1)]=E[(h_e(t_0)-h(t_0))(h_e(t_1)-h(t_1)]=0$. Therefore, $h_e$ is a modulated white noise process:
\begin{equation}
	h_e(t) = \sigma[h_e(t)] \cdot \nu(t).
\end{equation}
Here $\nu(t)$ is a white noise with $E[\nu(t)]=0,\sigma[\nu(t)]=1$ and $\forall t_0\neq t_1, \text{Cov}[\nu(t_0),\nu(t_1)]=0$.

\begin{figure}
	\centering
	\begin{tikzpicture}
		\begin{axis}[
			width = 0.47\textwidth,
			height = 0.3\textwidth,
			axis y line = left,
			axis x line = bottom,
			xlabel      = {$\omega$ / Hz},
			ylabel      = {Gain / dB},
			xmode = log,
			xmin = 20, xmax = 20000,
			ymin = -55, ymax = -15,
			legend cell align = left,
			legend pos = north east,
			legend transposed = true,
			legend style = {draw=none, fill=none},
			grid = major,
			/pgf/number format/.cd,
			use comma,
			1000 sep={}
		]
			\addplot[green-base-shiny, unbounded coords=jump] table [x=freq, y=IR, col sep=comma] {img/error_spectrum_IR.csv};
			\addlegendentry{$|\mathcal{F}(h)|^2$};
			\addplot[blue-base, unbounded coords=jump] table [x=freq, y=IR_var, col sep=comma] {img/error_spectrum_IR.csv};
			\addlegendentry{$\sigma^2[\mathcal{F}(h_e)]$};
			\addplot[red-base, unbounded coords=jump] plot coordinates {(20, -48.9712) (20000, -48.9712)};
			\addlegendentry{$\frac{1}{s}\mathcal{F}(\sigma^2[h_e])(0)$};
		\end{axis}
	\end{tikzpicture}
	\caption{Energy spectrum of IR and its noise on the audible range, produced by our simulator in the Sibenik scene (see \autoref{SEC:EXPERIMENT-STIMULI}). The deviation of IR variance from the estimation given by \autoref{eq:NoiseSpectrum} is about 5dB at low frequencies and less than 2dB at middle frequencies. However, this deviation is minimal compared to the IR energy spectrum itself, which is $|\mathcal{F}(h)|^2$.}
	\label{FIG:IRNOISEGAIN}
\end{figure}
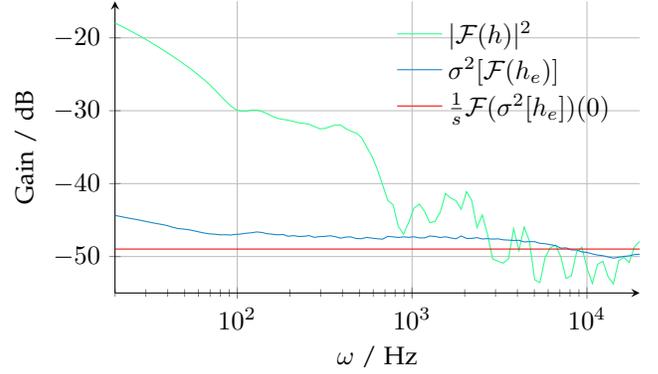

Now consider the Fourier spectrum $\mathcal{F}(h_e)$. We immediately know from $E[\nu(t)]=0$ that $E[\mathcal{F}(h_e)(\omega)]=0$, and
\begin{equation}
	\begin{aligned}
		& \text{Cov}[\mathcal{F}(h_e)(\omega_0), \mathcal{F}(h_e)(\omega_1)] \\
		=\ & E[\mathcal{F}(h_e)(\omega_0)\overline{\mathcal{F}(h_e)(\omega_1)}] \\
		=\ & \frac{1}{s^2}\sum_{t,u\in\mathbb{Z}}E[h_e(st)h_e(su)]e^{-2\pi is(\omega_0t-\omega_1u)} \\
		=\ & \frac{1}{s^2}\sum_{t\in\mathbb{Z}}\sigma^2[h_e(st)]e^{-2\pi i(\omega_0-\omega_1)st} \\
		=\ & \frac{1}{s}\mathcal{F}(\sigma^2[h_e])(\omega_0-\omega_1).
	\end{aligned}
\end{equation}
Specifically,
\begin{equation}
	\sigma^2[\mathcal{F}(h_e)(\omega)] = \frac{1}{s}\mathcal{F}(\sigma^2[h_e])(0) = \frac{1}{s^2}\sum_{t\in\mathbb{Z}}\sigma^2[h_e(st)].
	\label{eq:NoiseSpectrum}
\end{equation}
This is an interesting result: no matter what shape the IR noise envelope $\sigma^2[h_e]$ is, it is always equivalent to a white noise: the energy spectral density $\sigma^2[\mathcal{F}(h_e)(\omega)]$ is equal for all $\omega$. In actual path tracers, dependence between different parts of $h_e$ may exist, and the density spectrum may not be entirely flat. However, the fluctuation is largely negligible, especially when compared with the IR spectrum (see \autoref{FIG:IRNOISEGAIN}).

Given the input audio $f_i$, the energy spectral density of the output error $f_e$ is
\begin{equation}
    \label{eq:staticpowerspectrum}
    \begin{aligned}
        \sigma^2[\mathcal{F}(f_e)] &= \sigma^2[\mathcal{F}(f_i * h_e)] \\
        &=\sigma^2[\mathcal{F}(f_i)\mathcal{F}(h_e)] \\
        &=|\mathcal{F}(f_i)|^2\cdot\sigma^2[\mathcal{F}(h_e)].
    \end{aligned}
\end{equation}
Thus, we have the power spectrum estimation of the error signal in the static case.

Now we'll look at the relationship between the error power spectrum and IR energy SNR. We discretize the integral of SNR in \autoref{eq:SNRdefinition}, then use \autoref{eq:DiscreteParseval} and \autoref{eq:NoiseSpectrum}:
\begin{equation}
	\begin{aligned}
		\text{SNR}(h) &=\frac{\sum_{t\in\mathbb{Z}}h(st)^2}{E[\sum_{t\in\mathbb{Z}}h_e(st)^2]}\\
		&=\frac{\sum_{t\in\mathbb{Z}}h(st)^2}{\sum_{t\in\mathbb{Z}}\sigma^2[h_e(st)]}\\
		&=\frac{\int_{-s/2}^{s/2}|\mathcal{F}(h)(\omega)|^2d\omega}{s\cdot\sigma^2[\mathcal{F}(h_e)(\omega)]}.
	\end{aligned}
\end{equation}
Now with \autoref{eq:staticpowerspectrum}, we have
\begin{equation}
	\label{eq:SNRtoPowerSpectrum}
	\sigma^2[\mathcal{F}(f_e)] = |\mathcal{F}(f_i)|^2\cdot\frac{\int_{-s/2}^{s/2}|\mathcal{F}(h)(\omega)|^2d\omega}{s\cdot\text{SNR}(h)}.
\end{equation}
This is the relationship between the power spectrum of the error signal and the energy SNR of IR.

\subsection{Dynamic Case}

In this subsection, we discuss the case of real-time renderers, where the IR is recalculated in every frame (hence the name ``dynamic''). We assume that the scene remains unchanged during rendering. The IR difference between frames is thus only due to calculation error. We can regard the different IRs as different sample paths of the same variable $h_e$, the spectrum of which is given in the previous section. The resulting audio is produced by stitching multiple audio frames together. If we note the IR error of the $k$-th frame as $h_k$, then the output error signal is
\begin{equation}
	\label{eq:dynamicoutput}
	f_e(t) = \sum_{k\in\mathbb{Z}}\left((f_i * h_k)\cdot w(t-k\Delta t_w)\right),w(t)\in [0,1].
\end{equation}
Here $w(t)$ is the window function for frame interpolation and $\Delta t_w$ is the frame length. Let $g_k=f_i * h_k$, then $g_k$ is a sample path of $g = f_i * h_e$, and
\begin{equation}
	\mathcal{F}(f_e) = \sum_{k\in\mathbb{Z}}(\mathcal{F}(g_k) * \mathcal{F}(w)e^{2\pi i\omega k\Delta t_w}).
\end{equation}
Let $g^*_k = (f_i * h_k)\cdot w(t-k\Delta t_w)$, then
\begin{equation}
	\mathcal{F}(g^*_k)(\omega) = \int_{\mathbb{R}}\mathcal{F}(g)(x)\mathcal{F}(w)(\omega-x)e^{2\pi i(\omega-x) k\Delta t_w}dx,
\end{equation}
and
\begin{equation}
	\begin{aligned}
		\sigma^2[\mathcal{F}(g^*_k)(\omega)] = & \int_{\mathbb{R}^2}\Big(\text{Cov}[\mathcal{F}(g)(x),\mathcal{F}(g)(y)]\cdot \\
		& \qquad \mathcal{F}(w)(\omega-x)\mathcal{F}(w)(y-\omega)\cdot \\
		& \qquad e^{2\pi i(y-x) k\Delta t_w}\Big)dxdy.
	\end{aligned}
\end{equation}
Substituting $y-x$ with $y$, we have
\begin{equation}
	\begin{aligned}
		\sigma^2[\mathcal{F}(g^*_k)(\omega)] = &\int_{\mathbb{R}^2}\Big(\text{Cov}[\mathcal{F}(g)(x),\mathcal{F}(g)(x+y)]\cdot \\
		& \qquad \mathcal{F}(w)(\omega-x)\mathcal{F}(w)(x+y-\omega)\cdot \\
		& \qquad e^{2\pi iy k\Delta t_w}\Big)dxdy,
	\end{aligned}
\end{equation}
and
\begin{equation}
	\begin{aligned}
		\sigma^2[\mathcal{F}(f_e)(\omega)] =&\sum_{k\in\mathbb{Z}}\sigma^2[\mathcal{F}(g^*_k)(\omega)] \\
		=&\int_{\mathbb{R}^2}\Big(\text{Cov}[\mathcal{F}(g)(x),\mathcal{F}(g)(x+y)]\cdot \\
		& \qquad \mathcal{F}(w)(\omega-x)\mathcal{F}(w)(x+y-\omega)\cdot \\
		& \qquad \sum_{k\in\mathbb{Z}}e^{2\pi iy k\Delta t_w}\Big)dxdy.
	\end{aligned}
\end{equation}
The sum $\sum_{k\in\mathbb{Z}}e^{2\pi iy k\Delta t_w}$ is equal to $\Sha(\Delta t_w y)$, where $\Sha(x)$ is the shah function \cite{combfunction}:
\begin{equation}
	\Sha(x)=\sum_{n\in\mathbb{Z}}\delta(x-n).
\end{equation}
Therefore,
\begin{equation}
	\label{eq:fullerrorspectrum}
	\begin{aligned}
	& \sigma^2[\mathcal{F}(f_e)(\omega)]= \\
	& \quad\frac{1}{\Delta t_w}\sum_{k\in\mathbb{Z}}\int_{\mathbb{R}}\Big(\text{Cov}[\mathcal{F}(g)(x),\mathcal{F}(g)(x+\frac{k}{\Delta t_w})]\cdot \\
	& \qquad\qquad\qquad \mathcal{F}(w)(\omega-x)\mathcal{F}(w)(x+\frac{k}{\Delta t_w}-\omega)\Big)dx.
	\end{aligned}
\end{equation}
For a narrow-band $w$,
\begin{equation}
	\label{eq:approxerrorspectrum}
	\sigma^2[\mathcal{F}(f_e)(\omega)]\approx\frac{\sigma^2[\mathcal{F}(g)] * |\mathcal{F}(w)|^2}{\Delta t_w}.
\end{equation}
Here $\sigma^2[\mathcal{F}(g)]$ is equal to $\sigma^2[\mathcal{F}(f_e)]$ in \autoref{eq:SNRtoPowerSpectrum}. Therefore, we also have the relationship between $\sigma^2[\mathcal{F}(f_e)]$ and $\text{SNR}(h)$ for the dynamic case. In practice, the audio quantization error will also affect the spectrum of $f_e$. The effect of quantization adds an additional term $\frac{q^2}{3}$, where $q$ is the quantization interval.

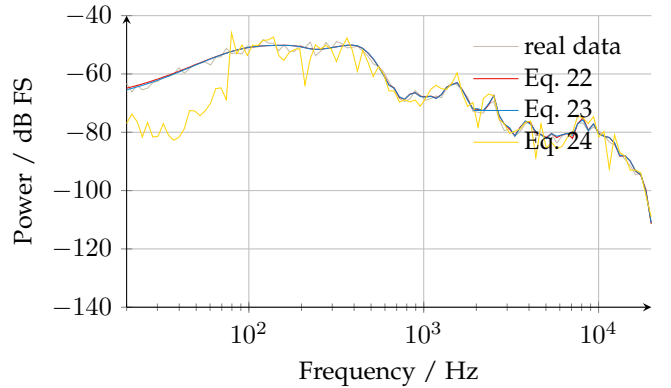
\begin{figure}[ht]
	\centering
	\begin{tikzpicture}
		\begin{axis}[
			width = 0.47\textwidth,
			height = 0.3\textwidth,
			axis y line = left,
			axis x line = bottom,
			xlabel      = {Frequency / Hz},
			ylabel      = {Power / dB FS},
			xmode = log,
			xmin = 20, xmax = 20000,
			ymin = -140, ymax = -40,
			legend cell align = left,
			legend pos = north east,
			legend transposed = true,
			legend style = {draw=none, fill=none},
			grid = major,
			/pgf/number format/.cd,
			use comma,
			1000 sep={}
		]
			\addplot[gray-base-thick, unbounded coords=jump] table [x=freq, y expr=ln(\thisrow{value}) * 10 / ln(10), col sep=comma] {img/spectrum_audio_10avg.csv};
			\addlegendentry{real data};
			\addplot[red-base, unbounded coords=jump] table [x=freq, y expr=ln(\thisrow{value}) * 10 / ln(10), col sep=comma] {img/spectrum_estimate1.csv};
			\addlegendentry{Eq.\ 22}; % \ref{fullerrorspectrum}
			\addplot[blue-base, unbounded coords=jump] table [x=freq, y expr=ln(\thisrow{value}) * 10 / ln(10), col sep=comma] {img/spectrum_estimate2.csv};
			\addlegendentry{Eq.\ 23}; % \ref{approxerrorspectrum}
			\addplot[yellow-base, unbounded coords=jump] table [x=freq, y expr=ln(\thisrow{value}) * 10 / ln(10), col sep=comma] {img/spectrum_estimate3.csv};
			\addlegendentry{Eq.\ 24}; % \ref{approxerrorspectrum2}
		\end{axis}
	\end{tikzpicture}
	\caption{Comparison of the noise signal spectrum generated by a dynamic sound rendering process and its estimations.}
	\label{FIG:NOISESPECTRUM}
\end{figure}

One may expect an even simpler estimation for $\sigma^2[\mathcal{F}(f_e)]$ if the bandwidth of $\omega$ is completely ignored. If we approximate $\mathcal{F}(w)$ with a Dirac delta function, we can replace \autoref{eq:approxerrorspectrum} with
\begin{equation}
	\label{eq:approxerrorspectrum2}
	\sigma^2[\mathcal{F}(f_e)(\omega)]\approx\sigma^2[\mathcal{F}(g)]\cdot\frac{\int_\mathbb{R}w(t)^2dt}{\Delta t_w}.
\end{equation}
However, this turns out to be an oversimplification. \autoref{FIG:NOISESPECTRUM} shows the results of various estimation formulae applied to a simulated dynamic rendering process. We can see that both \autoref{eq:fullerrorspectrum} and \autoref{eq:approxerrorspectrum} fit the real data very well, while \autoref{eq:approxerrorspectrum2} fails to predict the noise power at low frequencies. This is where the dynamic case differs from the static case --- the spectral distortion brought by the window function $w$ cannot be ignored.

\section{Quality Criterion}
\label{SEC:QUALITY-CRITERION}

Since our quality criterion is developed based on Zwicker's loudness, we need to first investigate Zwicker's loudness algorithm. Zwicker proposed two different loudness algorithms \cite{ZwickerLoudness3}, one for ``stationary'' sounds and another for ``time-varying'' sounds. The procedures of the two algorithms are largely similar. In this section, we assume the input audio clips to be stationary. That is, the spectral characteristics of the input audios do not change significantly in their durations.

\begin{figure*}[htbp]
	\centering
	\tikzstyle{data} = [rectangle, minimum width=2, minimum height=1, draw=black]
	\tikzstyle{proc} = [rectangle, rounded corners, minimum width=2, minimum height=2, draw=black, fill=gray-base]
	\tikzstyle{diff} = [fill=green-base-shiny]

	\subfloat[Zwicker's loudness]{
		\begin{tikzpicture}
			\node(signal)[data, yshift=15]{$f$};
			\node(label1)[above of=signal]{audio};
			\node(signalpwr)[data,right of=signal, xshift=50]{$p$};
			\node(label2)[above of=signalpwr]{\shortstack{power\\(28 bands)}};
			\draw[-stealth,line width=2pt](signal)--(signalpwr);
			\node(signalloudness)[data,right of=signalpwr, xshift=120]{$N$};
			\node(maskloudness)[data,below of=signalloudness, yshift=-15]{$N_t$};
			\node(label3)[above of=signalloudness]{\shortstack{core loudness\\(20 bands)}};
			\draw[-stealth,line width=2pt](signalpwr)--(signalloudness);

			\node(filters)[proc, right of=signal, xshift=10, yshift=-22, minimum width=40, minimum height=60]{\shortstack{band\\filters}};
			\node(correct1)[proc, right of=filters, xshift=55, minimum width=50, minimum height=60]{\shortstack{low freq\\correction\\1}};
			\node(stevenslaw)[proc, right of=correct1, xshift=30, minimum width=60, minimum height=60]{\shortstack{Steven's law}};
			\node(diff)[proc, right of=stevenslaw, xshift=18]{$\Delta$};
			\node(loudnesscurve)[data,right of=diff, xshift=110]{$l$};
			\draw[-stealth,line width=2pt](maskloudness)--(diff);
			\draw[-stealth,line width=2pt](signalloudness)--(diff);
			\draw[-stealth,line width=2pt](diff)--(loudnesscurve);
			\node(correct2)[proc, right of=stevenslaw, xshift=60, minimum width=50, minimum height=60]{\shortstack{low freq\\correction\\2}};
			\node(auditoryfilter)[proc, right of=correct2, xshift=25, minimum width=50, minimum height=60]{\shortstack{auditory\\filters}};
			\node(result)[data, right of=loudnesscurve, xshift=35]{$S$};
			\draw[-stealth,line width=2pt](loudnesscurve)--(result);

			\node(label3)[above of=loudnesscurve, yshift=22]{specific loudness};
			\node(label4)[above of=result, yshift=23]{result};
			\node(integral)[proc, right of=auditoryfilter, xshift=45, minimum width=20, minimum height=20]{$\int dx$};
		\end{tikzpicture}
	}\\
	\subfloat[proposed quality criterion]{
		\begin{tikzpicture}
			\node(mask)[data]{$f_m$};
			\node(signal)[data,above of=mask, yshift=15]{$f$};
			\node(label1)[above of=signal]{audio};
			\node(maskpwr)[data,right of=mask, xshift=50]{$p_m$};
			\node(signalpwr)[data,right of=signal, xshift=50]{$p$};
			\node(label2)[above of=signalpwr]{\shortstack{power\\(28 bands)}};
			\draw[-stealth,line width=2pt](mask)--(maskpwr);
			\draw[-stealth,line width=2pt](signal)--(signalpwr);
			\node(maskloudness)[data,right of=maskpwr, xshift=175]{$N_m$};
			\node(signalloudness)[data,right of=signalpwr, xshift=175]{$N$};
			\node(label3)[above of=signalloudness]{\shortstack{core loudness\\(20 bands)}};
			\draw[-stealth,line width=2pt](maskpwr)--(maskloudness);
			\draw[-stealth,line width=2pt](signalpwr)--(signalloudness);
			\node(maskloudnesscurve)[data,right of=maskloudness, xshift=55]{$l_m$};
			\node(signalloudnesscurve)[data,right of=signalloudness, xshift=55]{$l$};
			\node(label3)[above of=signalloudnesscurve]{specific loudness};
			\draw[-stealth,line width=2pt](maskloudness)--(maskloudnesscurve);
			\draw[-stealth,line width=2pt](signalloudness)--(signalloudnesscurve);

			\node(filters)[proc, right of=signal, xshift=10, yshift=-22, minimum width=40, minimum height=60]{\shortstack{band\\filters}};
			\node(correct1)[proc, right of=filters, xshift=55, minimum width=50, minimum height=60]{\shortstack{low freq\\correction\\1}};
			\node(stevenslaw)[proc, right of=correct1, xshift=30, minimum width=60, minimum height=60]{\shortstack{Steven's law}};
			\node(correct2)[proc, diff, right of=stevenslaw, xshift=30, minimum width=50, minimum height=60]{\shortstack{low freq\\correction\\2}};
			\node(auditoryfilter)[proc, diff, right of=correct2, xshift=55, minimum width=50, minimum height=60]{\shortstack{auditory\\filters}};
			\node(diff)[proc, diff, right of=auditoryfilter, xshift=15]{$\Delta$};
			\node(result)[data, right of=diff, xshift=35]{$S$};
			\draw[-stealth,line width=2pt](maskloudnesscurve)--(diff);
			\draw[-stealth,line width=2pt](signalloudnesscurve)--(diff);
			\draw[-stealth,line width=2pt](diff)--(result);
			\node(label4)[above of=result, yshift=23]{result};

			\node(integral)[proc, right of=auditoryfilter, xshift=45, minimum width=20, minimum height=20]{$\int dx$};
		\end{tikzpicture}
	}
	\caption{Diagrams for algorithms of Zwicker's loudness and our quality criterion. We use $\Delta$ and $\int dx$ to represent difference and integral operations. Green blocks signify operations different from Zwicker's original algorithm.}
	\label{FIG:LOUDNESSDIAGRAM}
\end{figure*}
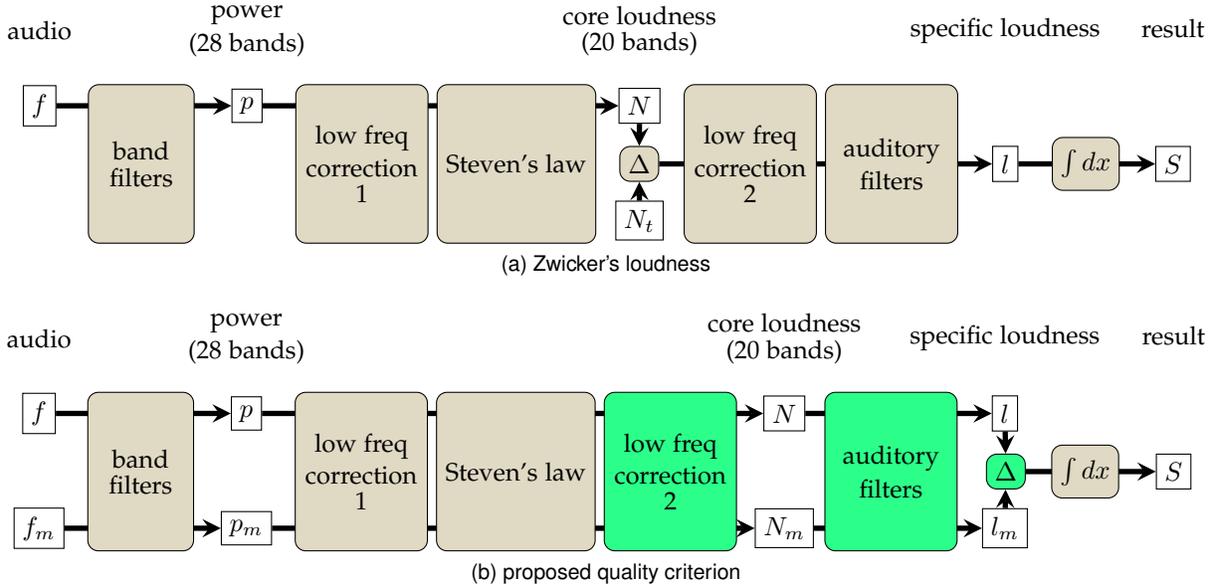

A simple diagram for the loudness algorithm for stationary sounds is given in \autoref{FIG:LOUDNESSDIAGRAM}. In the first step, the input audio $f$ is filtered by 28 band filters, and the power of each filtered audio is calculated. The results are expressed by a vector $p$. After this, $p$ will be converted into the specific loudness $l$ through a series of processes. $l$ is a function over the Bark scale, which maps the audible frequency range into a number between 0 and 24. The range $[n-1,n]$ on the scale is called the $n$-th critical band. The mapping scheme of the Bark scale is designed to make different critical bands approximately independent from each other in auditory perception. In the end, function $l$ is integrated over the Bark scale to achieve the total loudness, which is Zwicker's loudness. Refer to \cite{ZwickerLoudness3} for the detailed implementation of Zwicker's algorithm. 

One can see from \autoref{FIG:LOUDNESSDIAGRAM} that our criterion requires two input audio clips instead of one as in Zwicker's algorithm. Zwicker's algorithm takes an audio $f$ as input and outputs its loudness $S$ relative to the internal noise. Our criterion requires a masker $f_m$ and a full audio $f$, which is the sum of the masker and the masked signal. In the context of simulation quality assessment, $f_m=f_o$, $f=f_o + f_e$, where $f_o$ and $f_e$ are the ideal output and the error signal, respectively. We have already discussed the power spectrum of $f_o$ and $f_e$ in the previous section. One can use this information to calculate the value of $p$ and $p_m$ on different bands.

In Zwicker's algorithm, the loudness $N$ of each critical band (core loudness) is an important intermediate variable. This variable is calculated using Steven's power law \cite{Stevenslaw}, which gives the relationship between audio power $p$ and sensation magnitude \cite{ZwickerLoudness1}:
\begin{equation}
	N(p) = Cp^a
\label{eq:Stevenslaw}
\end{equation}
where $C$ and $a$ are parameters. Zwicker uses $C=0.03175\sqrt2$ and $a=0.25$ in his algorithm. The loudness of a certain band is the difference between the actual sensation magnitude and the sensation threshold:
\begin{equation}
	\Delta N = \max\{N(p+p_m)-N_t,0\},N_t=N(p_t+p_m)
\label{eq:ZwickerCoreLoudness}
\end{equation}
where $p$, $p_m$ and $p_t$ are power of input audio, internal noise and hearing threshold, respectively. Zwicker used $p_t = \frac{1}{3}p_m$ in his algorithm. At low frequencies, the above equation no longer reflects the actual human perception correctly, so Zwicker introduced a few nonlinear corrections into the algorithm to mitigate this problem.

With the core loudness known, we can calculate specific loudness by applying acoustic filters. The specific loudness curve is calculated as follows:
\begin{equation}
	l(x) = \max_{1\leq i\leq 20}\{\text{AF}_i(\Delta N_i)(x)\}
\label{eq:ZwickerSpecificLoudness}
\end{equation}
where $\Delta N_i$ is the core loudness of the $i$-th band and $\text{AF}_i(\Delta N_i)$ is the corresponding specific loudness curve. The value of $\text{AF}_i$ is equal to $\Delta N_i$ inside the i-th band, but also extends to higher frequencies.

The main difference between our criterion and Zwicker's loudness algorithm is that the mask is no longer predetermined. In our criterion, $p_m$ is the power of the input mask audio $f_m$ plus the internal noise power, and $p_t$ is no longer a static value. This leads to the possibility of upward spreading: When the mask audio is intensive on a certain band, its masking effect will also extend to higher frequencies like the function $\text{AF}_i$. Because of this, we calculate the difference of specific loudness instead of core loudness in our criterion.

From \autoref{FIG:LOUDNESSDIAGRAM}, we can see that the position change of the difference operation has some effect on two other operations. ``low frequency correction 2'' multiplies the core loudness of the first critical band by a factor $c$:
\begin{equation}
	c = \min\{0.4+0.32\Delta N^{0.2},1\}.
\end{equation}
To achieve a similar effect, we replace $c\Delta N$ with $c(N-N_t)+N_t$ and $c(N_m-N_t)+N_t$ in our criterion, where $N_t$ is the threshold in Zwicker's algorithm. For auditory filters, we use a linear approximation to replace $\text{AF}_i(\Delta N_i)$ in \autoref{eq:ZwickerSpecificLoudness}:
\begin{equation}
	\text{AF}_i(N)\approx \frac{N}{30}\cdot \text{AF}_i(30).
	\label{eq:AFLinearApprox}
\end{equation}
Using this approximation, we have $\text{AF}_i(N-N')\approx \text{AF}_i(N)-\text{AF}_i(N')$, making the difference operation and the application of auditory filters exchangeable. Fortunately, the original $\text{AF}_i$ is already very close to a linear function, and this approximation does not introduce too much error.

\subsection{Soft Threshold}
\label{SEC:SOFT-THRESHOLD}

While the relationship $p_t=\frac{1}{3}p_m$ is used in Zwicker's algorithm, experiments have shown that the ratio $k_p=p_t/p_m$ is not a fixed value but is highly related to frequency and relatively independent from $p_m$ \cite{lowfreqKvalue}. Therefore, we can model $k_p$ as a function of frequency. However, the function $k$ may also vary for different individuals, and we should take this variance into account if possible.

Our solution to the abovementioned problem is to use a ``soft threshold'' in our criterion. Suppose that we calculate $\Delta a$ with the equation below:
\begin{equation}
	\Delta a = \max\{a-a_t,0\}.
\end{equation}
The key point of the soft threshold is to regard $a_t$ as a distribution instead of a fixed value. we consider $\Delta a$ as a function of $a$, then we have
\begin{equation}
	E[\Delta a](a) = \int_{-\infty}^a\text{CDF}(a_t)dx.
\end{equation}
Here $\text{CDF}(a_t)$ is the cumulative distribution function of $a_t$. In our case, we assume that $a_t$ conforms to a logistic distribution. When $E[a_t]=E_a$ and $\sigma[a_t]=\sigma_a$, we have
\begin{equation}
	E[\Delta a] = s\ln(e^\frac{a-E_a}{s}+1),s=\frac{\sqrt{3}\sigma_a}{\pi}.
\end{equation}
We will take $E[\Delta a]$ as the difference under the soft threshold.

In our algorithm, we are calculating the difference of specific loudness $\Delta l$ instead of $\Delta N$ or $\Delta p$. Like Zwicker, we assume that $k_l=l_t/l_m$ is independent from $l_m$. The values of $E[k_l]$ and $\sigma[k_l]$ are measured by experiments.

To compare $k_l$ with $k_p$ values measured in previous papers, we ignore the low-frequency correction and derive the relationship of $p$ and $l$ from \autoref{eq:Stevenslaw} and \autoref{eq:AFLinearApprox}:
\begin{equation}
	l \propto C((p+p_m)^a-p_m^a).
\end{equation}
Thus, we have the approximate conversion formula
\begin{equation}
	k_l=C((1+k_p)^a-1).
	\label{eq:KConversion}
\end{equation}
% For loudness, $a$ is measured to be 0.3 in \cite{Stevenslaw} and 0.23 in \cite{ZwickerLoudness1}. 

\section{Experiments and Results}

In this section, we will show the results of our experiments. We first measured the values of $E[k_l]$ and $\sigma[k_l]$ and tested the independence between $k_l$ and $l_m$. Then we test our criterion on real outputs. We achieve these goals by measuring the audible power threshold of different signals under masking. The details are presented below.

\subsection{Experiment Procedure}

\subsubsection{Stimuli}

\label{SEC:EXPERIMENT-STIMULI}

In our experiment, we ask the subjects to listen to different audio clips with Beyerdynamic DT770 headphones in a quiet room. The background noise of the experiment environment is 30-45 dB SPL. The audio clips are played by a web application on a regular PC. All the audio clips are generated and played in the 16-bit, 48kHz stereo format. The contents of the left and right channels are identical.

When measuring $k_l$, we use sinusoidal waves of different frequencies as signals and pink noise as maskers. The power spectrum of a pink noise signal is inversely proportional to frequency, which is similar to many natural sounds. In our experiment, we use pink noise generated by the stochastic Voss-McCartney algorithm \cite{VossPinkNoise} at 55 dB SPL. The frequencies of sinusoidal waves are the midpoints of the 24 critical bands defined in \cite{psychoacousticsbook}, which is different from the specification in \cite{ZwickerLoudness3}. We change the definition of the first band from 0-100Hz to 20-100Hz in our experiment. Each band corresponds to a test case, and both are indexed by the upper bound Bark value of the bands. This gives 24 test cases in total.

To validate the independence between $k_l$ and $l_m$, we measured the audible threshold of sinusoidal signals on bands 1, 7, 9, and 19 with maskers of varied intensities. We used 6 different masker power levels equally spaced between 50 and 70 dB SPL. This validation process also requires 24 test cases.

When testing our criterion on real outputs, we generate the audio clips similar to real sound renderers. First, we simulate sound propagation using path-tracing-based simulators in different scenes. Here we tried to cover all the ``typical varieties'' of sound environments. The reverberation length and the presence of direct contribution (DC) are two of the most prominent factors that affect user perception. Thus we used 4 scene configurations with 2 different geometry models (see \autoref{FIG:FLOORPLANS}). The configurations are listed below:

\begin{figure}
	\centering
	\subfloat[Sibenik]{
		\includegraphics[width=0.235\textwidth]{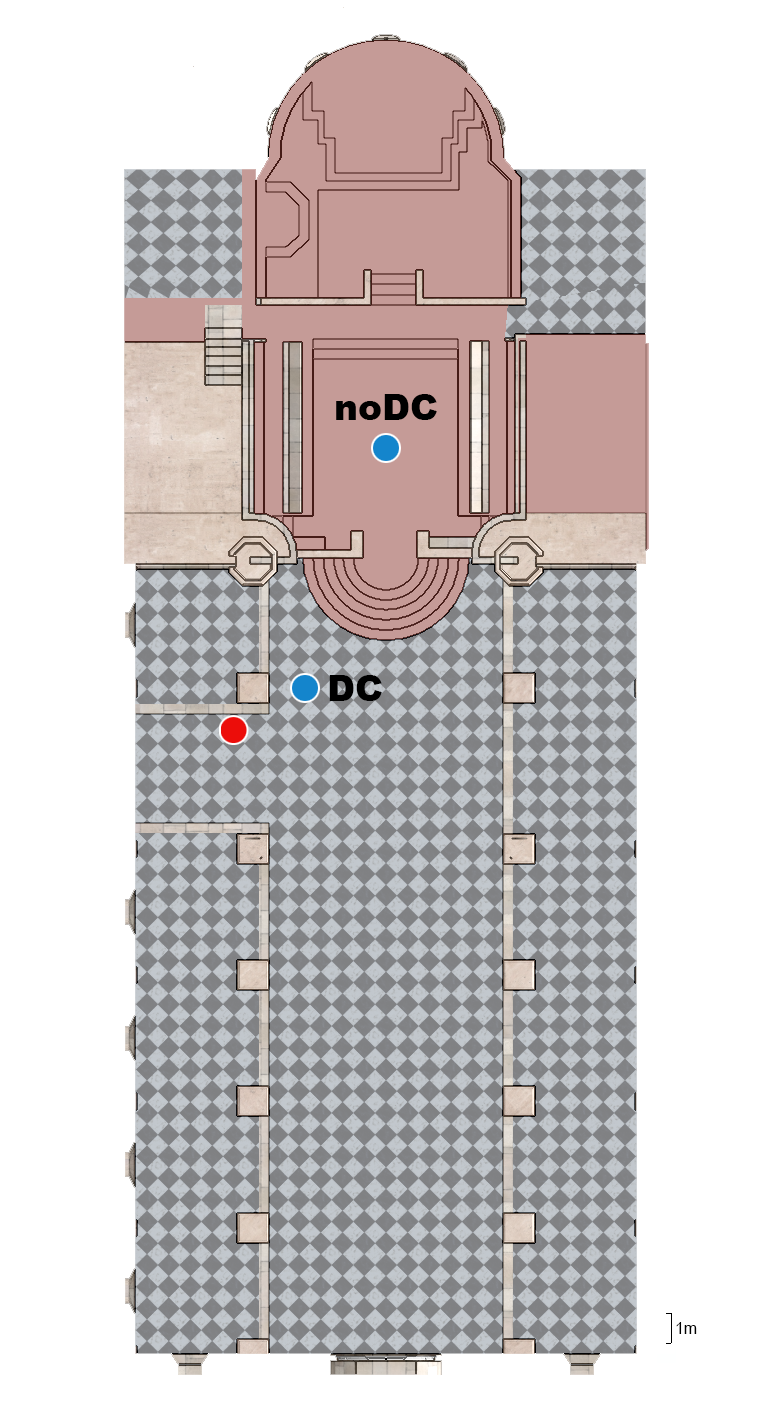}
	}
	\subfloat[Roomset]{
		\includegraphics[width=0.235\textwidth]{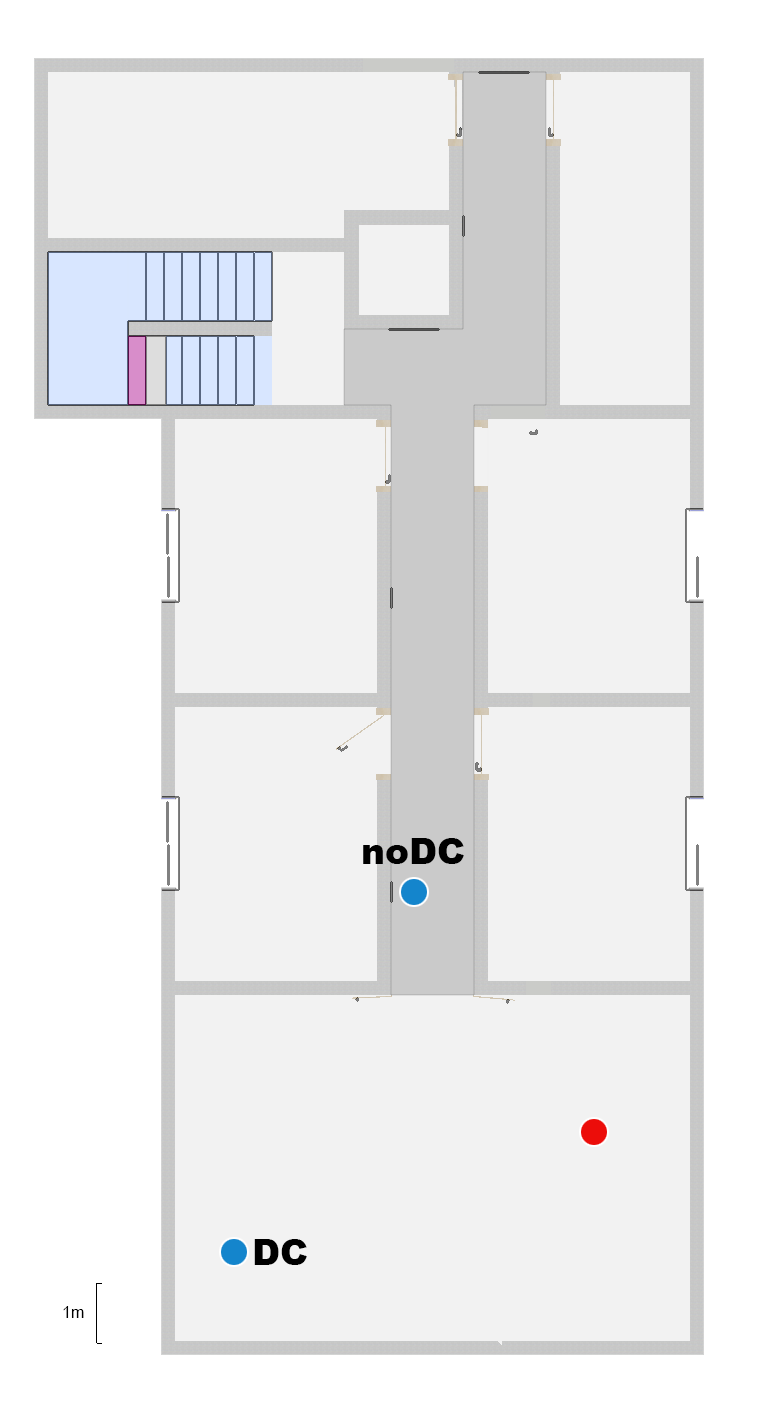}
	}
	\caption{Floor plans of the geometry models used in our experiment. The position of the sound source and listener are marked with red and blue dots.}
	\label{FIG:FLOORPLANS}
\end{figure}

\begin{itemize}
	\item Sibenik-DC: Interior model of a cathedral with a long reverberation time. The audio source and the listener are mutually visible so that the direct contribution (DC) component is present in the resultant IRs.
	\item Sibenik-noDC: Same as above, but now the source is not visible from the listener position.
	\item Roomset-DC: Interior model of a two-story guesthouse with multiple small rooms and narrow corridors. The reverberation time is relatively short. The source is visible from the listener position.
	\item Roomset-noDC: Same as above. The source is now invisible.
\end{itemize}
We computed 2000 IRs in each configuration. Each IR is computed from 48000 sampled paths, and the maximum number of reflections in a path is 40. We take the average of the 2000 IRs as the ``reference IR'' and the difference between each IR and the reference IR as its numerical error.

To validate the generalizability of our criterion, we used two different simulators for IR generation. One of the simulators uses the bidirectional path tracing (BDPT) algorithm \cite{cao2016BDPT}, and the other one uses the hybrid image source (HIS) method \cite{rindel1995computer}, where we use image sources generated with forward path tracing. Note that the HIS method doesn't violate the independence assumption in \autoref{section:static-case}, and our analysis is still valid.

After IR generation, we generate the masker audio clip by convoluting the input audio with the reference IR. We scaled the masker audio so that its power level is around 65 dB SPL when being played. The signal clip is generated by convoluting the input with IR errors according to \autoref{eq:dynamicoutput}. Here we set $w$ to a 512-sample Hanning window and $\Delta t_w=256$. The IR error of each frame is randomly chosen from the 2000 candidates.

We used 6 different input audio clips in our experiment:
\begin{itemize}
	\item bass: A music piece played by synthetic bass. The power spectrum concentrates at around 70Hz.
	\item piano: A piano piece whose energy distributes broadly in the 350-1000Hz range.
	\item voice: Speech of a male narrator. The energy concentrates on the 60-600Hz range.
	\item drums: Intense outdoor percussion music. The energy distributes broadly on the 100-2500Hz range.
	\item drips: Sound of dripping water. The energy distributes mainly between 600-3000Hz.
	\item nature: Environmental noise recorded on the Patagonian plain with rain sounds and occasional animal callings. The power spectrum is very similar to pink noise.
\end{itemize}
The above list covers most types of input audios that may appear in virtual acoustic simulations. We consider the first three audio clips as ``melodic'' because they contain pitch-related semantic information, which is usually sensitive to spectrum distortion. The last three audio clips are considered ``noisy'' because of the absence of such information. The combination of 6 input audio clips and 4 scene configurations gives 24 test cases in total.

After generation, we crop all the signal and masker clips to the first 10 seconds and apply an exponential fading to the beginning and the end of each clip. The fading duration is 0.5s.

% In the result section below, we will refer to our masking test configurations by the signal and the critical band index. For example, the configuration ``sinusoidal band 1'' indicates that the signal is a 60Hz sinusoidal wave and the masker is pink noise. The identity filter is referred as ``full-band''.

\subsubsection{Method}

We use the following trial to determine if a masked signal is audible to a subject. We present 4 candidate clips to the subject and the masker clip for reference. 1-3 of these candidates are the masked signal clip, and the rest are masker only, leading to 14 different possible combinations. The probability for each combination is the same. We require the subject to listen to the candidates and select all the clips containing the signal or press ``give up'' if they cannot tell a difference. All the candidate clips are replayable until the subject has submitted their answer. If the answer is correct, the signal is considered audible under masking.

Given a masker, the audible power threshold of a signal is determined by a bisection search. We set the initial width of the search range to 60 dB on the logarithmic scale and choose the upper bound empirically. For sinusoidal signals masked by pink noise, the upper bound is 30-40 dB above the power of the masker. We set the signal power to the midpoint of the search range and test its audibility. The midpoint is used as the new upper bound if the masked signal is audible or the new lower bound otherwise. We repeat this process until the width of the search range is below 1 dB and take the midpoint as the measured threshold.

% To accommodate for hearing difference between individuals, we measure the sound intensity with sensation level (dB SL) instead of sound pressure level (dB SPL). For each subject, we use the same bisection method to measure the non-masked hearing threshold of 1kHz sinusoidal signal. This threshold is used as the reference level (3 dB SL) of the subject. In calculation, we obtain the pressure level using the conversion formula $0$ dB SL $= 1.5$ dB SPL. This relationship is implied in \cite{ZwickerLoudness3}. 

\subsubsection{Subjects}

Our experiments involve 47 subjects in total. These subjects are divided into two age groups. The first group includes all subjects aged 18-27 and contains 37 subjects (16 female and 21 male). The age group is similar to the one for measuring the absolute threshold of hearing used in Zwicker's original algorithm \cite{psychoacousticsbook}. To examine the subjects' hearing condition, We performed pure tone audiometry tests in the range of 250-8000 Hz. For every subject, the measured absolute hearing threshold is less than 25 dB HL on the whole frequency range, which indicates normal hearing. The second group contains 10 subjects (5 female and 5 male) aged 28-55. According to audiometry results, 2 out of 10 of these subjects have their absolute hearing threshold in the range of 25-40 dB HL on certain frequencies, which indicates mild hearing impairment.

We divide the test cases into 3 groups according to their purposes (measurement of $k_l$, validation of independence, validation on real outputs). For $k_l$ measurement and independence validation, we allocate 10 subjects of age 18-27 to both test groups. This subject group size is the same as the prior work \cite{lowfreqKvalue}, which measures $k_l$ at low frequencies. For validation on real outputs, the allocation scheme will be detailed in \autoref{section:real-outputs}. In each test group, we measure the audible thresholds of test subjects under all 24 cases, which takes 1-2 hours to finish.

% subjects with impaired hearing: meihua yu and jinxing zhu

None of the subjects has previous experience in psychoacoustic experiments.

\subsection{Result}

\subsubsection{Measurement of $k_l$}

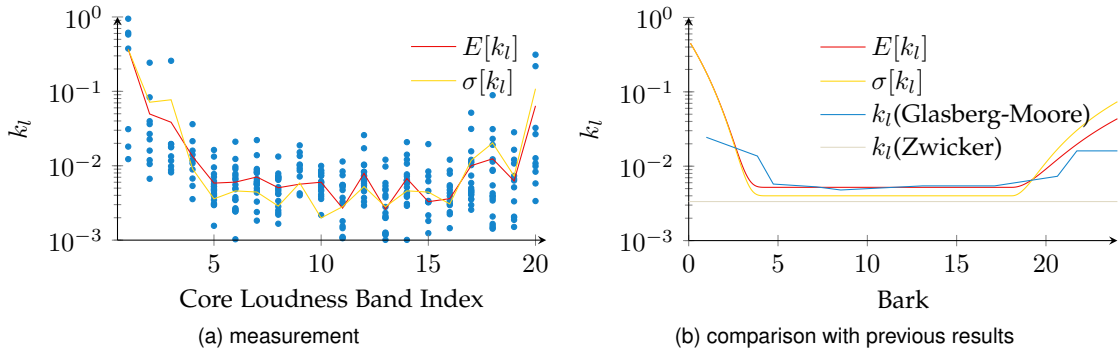
\begin{figure*}[htbp]
	\centering
	\subfloat[measurement]{
		\begin{tikzpicture}
			\begin{axis}[
				width = 0.4\textwidth,
				height = 0.25\textwidth,
				axis y line = left,
				axis x line = bottom,
				xlabel      = {Core Loudness Band Index},
				ylabel      = {$k_l$},
				xmin = 0.5, xmax = 20.5,
				ymin = 0.001, ymax = 1,
				ymode = log,
				legend cell align = left,
				legend pos = north east,
				legend transposed = true,
				legend style = {draw=none, fill=none},
				set layers,
				mark layer=axis background,
				/pgf/number format/.cd,
				1000 sep={}
			]
				\addplot[red-base, unbounded coords=jump] table [x=index, y=mean, col sep=comma] {img/k_statistics.csv};
				\addlegendentry{$E[k_l]$};
				\addplot[yellow-base, unbounded coords=jump] table [x=index, y=std, col sep=comma] {img/k_statistics.csv};
				\addlegendentry{$\sigma[k_l]$};
				\addplot[blue-base, mark=*, mark size=1pt, mark layer=axis background, only marks] table [x=index, y=k, col sep=comma] {img/k_values.csv};
				\addplot[blue-base, mark=*, mark size=1pt, mark layer=axis background, only marks] table [x=index, y=k, col sep=comma] {img/k_values_outlier.csv};
			\end{axis}
		\end{tikzpicture}
		\label{FIG:K-MEASUREMENT}
	}
	\subfloat[comparison with previous results]{
		\begin{tikzpicture}
			\begin{axis}[
				width = 0.4\textwidth,
				height = 0.25\textwidth,
				axis y line = left,
				axis x line = bottom,
				xlabel      = {Bark},
				ylabel      = {$k_l$},
				xmin = 0, xmax = 24,
				ymin = 0.001, ymax = 1,
				ymode = log,
				legend cell align = left,
				legend pos = north east,
				legend transposed = true,
				legend style = {draw=none, fill=none},
				set layers,
				mark layer=axis background,
				/pgf/number format/.cd,
				1000 sep={}
			]
				\addplot[red-base, unbounded coords=jump] table [x=bark, y=mean, col sep=comma] {img/k_mean_std.csv};
				\addlegendentry{$E[k_l]$};
				\addplot[yellow-base, unbounded coords=jump] table [x=bark, y=std, col sep=comma] {img/k_mean_std.csv};
				\addlegendentry{$\sigma[k_l]$};
				\addplot[blue-base, unbounded coords=jump] table [x=bark, y=k, col sep=comma] {img/k_GlasbergMoore.csv};
				\addlegendentry{$k_l$(Glasberg-Moore)};
				\addplot[gray-base, unbounded coords=jump] plot coordinates {(0, 0.003348) (24, 0.003348)};
				\addlegendentry{$k_l$(Zwicker)};
			\end{axis}
		\end{tikzpicture}
		\label{FIG:K-COMPARISON}
	}
	% \subfigure[relationship with $l_m$]{
	% 	\begin{tikzpicture}
	% 		\begin{axis}[
	% 			width = 0.3\textwidth,
	% 			height = 0.3\textwidth,
	% 			axis y line = left,
	% 			axis x line = bottom,
	% 			xlabel      = {$l_m$},
	% 			ylabel      = {$k_l$},
	% 			xmin = 0.08, xmax = 5,
	% 			ymin = 0.001, ymax = 2.5,
	% 			xmode = log,
	% 			ymode = log,
	% 			legend cell align = left,
	% 			legend pos = north east,
	% 			legend transposed = true,
	% 			legend style = {draw=none, fill=none},
	% 			set layers,
	% 			mark layer=axis background,
	% 			/pgf/number format/.cd,
	% 			1000 sep={}
	% 		]
	% 			\addplot[red-base, mark=*, mark size=1pt, mark layer=axis background, only marks] table [x=mask, y=k, restrict expr to domain={\thisrow{index}}{1:1}, col sep=comma] {img/k_trend.csv};
	% 			\addlegendentry{band 1};
	% 			\addplot[green-base, mark=*, mark size=1pt, mark layer=axis background, only marks] table [x=mask, y=k, restrict expr to domain={\thisrow{index}}{7:7}, col sep=comma] {img/k_trend.csv};
	% 			\addlegendentry{band 7};
	% 			\addplot[yellow-base, mark=*, mark size=1pt, mark layer=axis background, only marks] table [x=mask, y=k, restrict expr to domain={\thisrow{index}}{9:9}, col sep=comma] {img/k_trend.csv};
	% 			\addlegendentry{band 9};
	% 			\addplot[blue-base, mark=*, mark size=1pt, mark layer=axis background, only marks] table [x=mask, y=k, restrict expr to domain={\thisrow{index}}{19:19}, col sep=comma] {img/k_trend.csv};
	% 			\addlegendentry{band 19};
	% 		\end{axis}
	% 	\end{tikzpicture}
	% 	\label{FIG:K-DEPENDENCY}
	% }
	\caption{Visualization of various information about $k_l$. Figure (a) shows the values of $k_l$ on different core loudness bands measured from our experiment. Figure (b) shows the comparison of the fitting curve for $k_l$ given in \autoref{eq:EK} and \autoref{eq:SK} with previous results.}
	\label{FIG:K-INFO}
\end{figure*}

The influence of sinusoidal signals is always concentrated on the influence range of one core loudness band. Thus we can determine the value of $k_l$ on every single band from measured audibility thresholds. \autoref{FIG:K-MEASUREMENT} shows the value of measured $k_l$ from sinusoidal band 1-24 mask tests.

To use the measured data in our criterion, we need to fit $E[k_l]$ and $\sigma[k_l]$ with functions on the Bark scale. In the actual criterion, we use the following piecewise polynomials:
\begin{equation}
	E[k_l] =\left\{\begin{array}{ll}
		0.0013(4.4-x)^4 + 0.0052, & x < 4.4 \\
		0.0052, & 4.4\leq x\leq 18.1 \\
		0.0011(x-18.1)^2 + 0.0052, & x > 18.1
	\end{array}\right.
	\label{eq:EK}
\end{equation}
\begin{equation}
	\sigma[k_l] =\left\{\begin{array}{ll}
		0.0013(4.4-x)^4 + 0.004, & x < 4.4 \\
		0.004, & 4.4\leq x\leq 18.1 \\
		0.002(x-18.1)^2 + 0.004, & x > 18.1
	\end{array}\right.
	\label{eq:SK}
\end{equation}

To compare with previous data, we convert $k_p$ values measured by Glasberg and Moore in \cite{ZwickerLoudness2} and $k_p=1/3$ in Zwicker's algorithm into $k_l$ curves on the bark scale. We use the conversion formula \autoref{eq:KConversion} and the frequency-to-Bark mapping function from \cite{BarkScaleZwicker1980}. The result is displayed in \autoref{FIG:K-COMPARISON}, with curves defined by \autoref{eq:EK} and \autoref{eq:SK}. We can see that \autoref{eq:EK} matches well with previously measured data at central frequencies and shows a similar trend at low and high frequencies.

For the independence assumption of $k_l$ from $l_m$, we calculated the Pearson correlation coefficient between $k_l$ and $l_m$ on four bands with measured $k_l$ values under varied $l_m$. The result is given in \autoref{TABLE:K-CORRELATION}. $r<0$ implies that the value of $k_l$ tends to decrease when the masker loudness increases. This trend is negligible for bands 7 and 9, but not for bands 1 and 19. Hence the independent assumption holds for middle frequencies but is not entirely valid for low and high frequencies. This will not affect the accuracy of our criterion in most cases, but it may have some effect when the input sound has a very intense low/high-frequency component. We will not fix this problem in the current criterion for the sake of simplicity.

\begin{table}
	\caption{Pearson correlation coefficient $r$ of $k_l$ and $l_m$ on different bands.}
	\begin{center}
		\begin{tabular}{ccccc}
			\toprule
			band index & 1 & 7 & 9 & 19 \\
			\midrule
			$r$ & -0.5084 & -0.0384 & -0.0298 & -0.2593 \\
			\bottomrule
		\end{tabular}
	\end{center}
	\label{TABLE:K-CORRELATION}
\end{table}

\subsubsection{Application on Real Outputs}
\label{section:real-outputs}

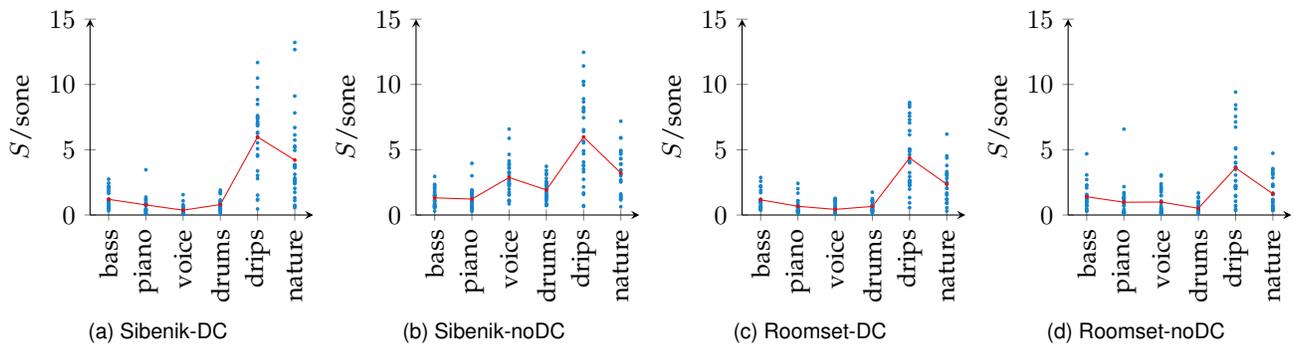
\begin{figure*}[htbp]
	\centering
	\subfloat[Sibenik-DC]{
		\begin{tikzpicture}
			\begin{axis}[
				width = 0.25\textwidth,
				height = 0.23\textwidth,
				axis y line = left,
				axis x line = bottom,
				xtick={1,2,3,4,5,6},
				xticklabels={bass,piano,voice,drums,drips,nature},
				x tick label style={rotate=90},
				ytick={},
				xlabel={},
				ylabel={$S$/sone},
				xmin = 0.5, xmax = 6.5,
				ymin = 0, ymax = 15,
				/pgf/number format/.cd
			]
			\addplot[blue-base, mark=*, mark size=0.5pt, mark layer=axis background, only marks] table [x=input,y=loudness_0, unbounded coords=discard, restrict expr to domain={(\thisrow{scene} - 1) * 2 + \thisrow{flag}}{1:1}, unbounded coords=discard, col sep=comma] {img/actual_sound_loudness_ag_lo.csv};
			% \addplot[gray-base, mark=*, mark size=0.5pt, mark layer=axis background, only marks] table [x=input,y=loudness_0, unbounded coords=discard, restrict expr to domain={(\thisrow{scene} - 1) * 2 + \thisrow{flag}}{0:0}, unbounded coords=discard, col sep=comma] {img/actual_sound_loudness_new.csv};
			\addplot[red-base, mark=*, mark size=0.5pt] table [x=input,y=avg, unbounded coords=discard, restrict expr to domain={(\thisrow{scene} - 1)}{0:0}, unbounded coords=discard, col sep=comma] {img/actual_sound_loudness_ag_lo_avg.csv};
			\end{axis}
		\end{tikzpicture}
	}
	\subfloat[Sibenik-noDC]{
		\begin{tikzpicture}
			\begin{axis}[
				width = 0.25\textwidth,
				height = 0.23\textwidth,
				axis y line = left,
				axis x line = bottom,
				xtick={1,2,3,4,5,6},
				xticklabels={bass,piano,voice,drums,drips,nature},
				x tick label style={rotate=90},
				ytick={},
				xlabel={},
				ylabel={$S$/sone},
				xmin = 0.5, xmax = 6.5,
				ymin = 0, ymax = 15,
				/pgf/number format/.cd
			]
			\addplot[blue-base, mark=*, mark size=0.5pt, mark layer=axis background, only marks] table [x=input,y=loudness_0, unbounded coords=discard, restrict expr to domain={(\thisrow{scene} - 2) * 2 + \thisrow{flag}}{1:1}, unbounded coords=discard, col sep=comma] {img/actual_sound_loudness_ag_lo.csv};
			% \addplot[gray-base, mark=*, mark size=0.5pt, mark layer=axis background, only marks] table [x=input,y=loudness_0, unbounded coords=discard, restrict expr to domain={(\thisrow{scene} - 2) * 2 + \thisrow{flag}}{0:0}, unbounded coords=discard, col sep=comma] {img/actual_sound_loudness_new.csv};
			\addplot[red-base, mark=*, mark size=0.5pt] table [x=input,y=avg, unbounded coords=discard, restrict expr to domain={(\thisrow{scene} - 2)}{0:0}, unbounded coords=discard, col sep=comma] {img/actual_sound_loudness_ag_lo_avg.csv};
			\end{axis}
		\end{tikzpicture}
	}
	\subfloat[Roomset-DC]{
		\begin{tikzpicture}
			\begin{axis}[
				width = 0.25\textwidth,
				height = 0.23\textwidth,
				axis y line = left,
				axis x line = bottom,
				xtick={1,2,3,4,5,6},
				xticklabels={bass,piano,voice,drums,drips,nature},
				x tick label style={rotate=90},
				ytick={},
				xlabel={},
				ylabel={$S$/sone},
				xmin = 0.5, xmax = 6.5,
				ymin = 0, ymax = 15,
				/pgf/number format/.cd
			]
			\addplot[blue-base, mark=*, mark size=0.5pt, mark layer=axis background, only marks] table [x=input,y=loudness_0, unbounded coords=discard, restrict expr to domain={(\thisrow{scene} - 3) * 2 + \thisrow{flag}}{1:1}, unbounded coords=discard, col sep=comma] {img/actual_sound_loudness_ag_lo.csv};
			% \addplot[gray-base, mark=*, mark size=0.5pt, mark layer=axis background, only marks] table [x=input,y=loudness_0, unbounded coords=discard, restrict expr to domain={(\thisrow{scene} - 3) * 2 + \thisrow{flag}}{0:0}, unbounded coords=discard, col sep=comma] {img/actual_sound_loudness_new.csv};
			\addplot[red-base, mark=*, mark size=0.5pt] table [x=input,y=avg, unbounded coords=discard, restrict expr to domain={(\thisrow{scene} - 3)}{0:0}, unbounded coords=discard, col sep=comma] {img/actual_sound_loudness_ag_lo_avg.csv};
			\end{axis}
		\end{tikzpicture}
	}
	\subfloat[Roomset-noDC]{
		\begin{tikzpicture}
			\begin{axis}[
				width = 0.25\textwidth,
				height = 0.23\textwidth,
				axis y line = left,
				axis x line = bottom,
				xtick={1,2,3,4,5,6},
				xticklabels={bass,piano,voice,drums,drips,nature},
				x tick label style={rotate=90},
				ytick={},
				xlabel={},
				ylabel={$S$/sone},
				xmin = 0.5, xmax = 6.5,
				ymin = 0, ymax = 15,
				/pgf/number format/.cd
			]
			\addplot[blue-base, mark=*, mark size=0.5pt, mark layer=axis background, only marks] table [x=input,y=loudness_0, unbounded coords=discard, restrict expr to domain={(\thisrow{scene} - 4) * 2 + \thisrow{flag}}{1:1}, unbounded coords=discard, col sep=comma] {img/actual_sound_loudness_ag_lo.csv};
			% \addplot[gray-base, mark=*, mark size=0.5pt, mark layer=axis background, only marks] table [x=input,y=loudness_0, unbounded coords=discard, restrict expr to domain={(\thisrow{scene} - 4) * 2 + \thisrow{flag}}{0:0}, unbounded coords=discard, col sep=comma] {img/actual_sound_loudness_new.csv};
			\addplot[red-base, mark=*, mark size=0.5pt] table [x=input,y=avg, unbounded coords=discard, restrict expr to domain={(\thisrow{scene} - 4)}{0:0}, unbounded coords=discard, col sep=comma] {img/actual_sound_loudness_ag_lo_avg.csv};
			\end{axis}
		\end{tikzpicture}
	}
	\caption{Error loudness at measured audibility thresholds in different test cases according to our criterion (age group 18-27). The audio is generated with the BDPT simulator. A low threshold value indicates that the error is easily perceptible by the listener. The measured threshold loudness values are marked with blue dots, and the average thresholds for different input audios are connected with red lines.}
	\label{FIG:THRESHOLD-BY-CASES}
\end{figure*}

\begin{figure*}[htbp]
	\centering
	\subfloat[Sibenik-DC]{
		\begin{tikzpicture}
			\begin{axis}[
				width = 0.25\textwidth,
				height = 0.23\textwidth,
				axis y line = left,
				axis x line = bottom,
				xtick={1,2,3,4,5,6},
				xticklabels={bass,piano,voice,drums,drips,nature},
				x tick label style={rotate=90},
				ytick={},
				xlabel={},
				ylabel={$S$/sone},
				xmin = 0.5, xmax = 6.5,
				ymin = 0, ymax = 15,
				/pgf/number format/.cd
			]
			\addplot[blue-base, mark=*, mark size=0.5pt, mark layer=axis background, only marks] table [x=input,y=loudness_0, unbounded coords=discard, restrict expr to domain={((\thisrow{scene} - 1) * 2 + (\thisrow{flag} - 1)) * 10 + \thisrow{tester}}{1:8}, unbounded coords=discard, col sep=comma] {img/actual_sound_loudness_ag_hi.csv};
			\addplot[gray-base-dark, mark=*, mark size=1pt, mark layer=axis background, only marks] table [x=input,y=loudness_0, unbounded coords=discard, restrict expr to domain={((\thisrow{scene} - 1) * 2 + (\thisrow{flag} - 1)) * 10 + \thisrow{tester}}{9:10}, unbounded coords=discard, col sep=comma] {img/actual_sound_loudness_ag_hi.csv};
			\addplot[red-base, mark=*, mark size=0.5pt] table [x=input,y=avg, unbounded coords=discard, restrict expr to domain={(\thisrow{scene} - 1)}{0:0}, unbounded coords=discard, col sep=comma] {img/actual_sound_loudness_ag_hi_avg.csv};
			\end{axis}
		\end{tikzpicture}
	}
	\subfloat[Sibenik-noDC]{
		\begin{tikzpicture}
			\begin{axis}[
				width = 0.25\textwidth,
				height = 0.23\textwidth,
				axis y line = left,
				axis x line = bottom,
				xtick={1,2,3,4,5,6},
				xticklabels={bass,piano,voice,drums,drips,nature},
				x tick label style={rotate=90},
				ytick={},
				xlabel={},
				ylabel={$S$/sone},
				xmin = 0.5, xmax = 6.5,
				ymin = 0, ymax = 15,
				/pgf/number format/.cd
			]
			\addplot[blue-base, mark=*, mark size=0.5pt, mark layer=axis background, only marks] table [x=input,y=loudness_0, unbounded coords=discard, restrict expr to domain={((\thisrow{scene} - 2) * 2 + (\thisrow{flag} - 1)) * 10 + \thisrow{tester}}{1:8}, unbounded coords=discard, col sep=comma] {img/actual_sound_loudness_ag_hi.csv};
			\addplot[gray-base-dark, mark=*, mark size=1pt, mark layer=axis background, only marks] table [x=input,y=loudness_0, unbounded coords=discard, restrict expr to domain={((\thisrow{scene} - 2) * 2 + (\thisrow{flag} - 1)) * 10 + \thisrow{tester}}{9:10}, unbounded coords=discard, col sep=comma] {img/actual_sound_loudness_ag_hi.csv};
			\addplot[red-base, mark=*, mark size=0.5pt] table [x=input,y=avg, unbounded coords=discard, restrict expr to domain={(\thisrow{scene} - 2)}{0:0}, unbounded coords=discard, col sep=comma] {img/actual_sound_loudness_ag_hi_avg.csv};
			\end{axis}
		\end{tikzpicture}
	}
	\subfloat[Roomset-DC]{
		\begin{tikzpicture}
			\begin{axis}[
				width = 0.25\textwidth,
				height = 0.23\textwidth,
				axis y line = left,
				axis x line = bottom,
				xtick={1,2,3,4,5,6},
				xticklabels={bass,piano,voice,drums,drips,nature},
				x tick label style={rotate=90},
				ytick={},
				xlabel={},
				ylabel={$S$/sone},
				xmin = 0.5, xmax = 6.5,
				ymin = 0, ymax = 15,
				/pgf/number format/.cd
			]
			\addplot[blue-base, mark=*, mark size=0.5pt, mark layer=axis background, only marks] table [x=input,y=loudness_0, unbounded coords=discard, restrict expr to domain={((\thisrow{scene} - 3) * 2 + (\thisrow{flag} - 1)) * 10 + \thisrow{tester}}{1:8}, unbounded coords=discard, col sep=comma] {img/actual_sound_loudness_ag_hi.csv};
			\addplot[gray-base-dark, mark=*, mark size=1pt, mark layer=axis background, only marks] table [x=input,y=loudness_0, unbounded coords=discard, restrict expr to domain={((\thisrow{scene} - 3) * 2 + (\thisrow{flag} - 1)) * 10 + \thisrow{tester}}{9:10}, unbounded coords=discard, col sep=comma] {img/actual_sound_loudness_ag_hi.csv};
			\addplot[red-base, mark=*, mark size=0.5pt] table [x=input,y=avg, unbounded coords=discard, restrict expr to domain={(\thisrow{scene} - 3)}{0:0}, unbounded coords=discard, col sep=comma] {img/actual_sound_loudness_ag_hi_avg.csv};
			\end{axis}
		\end{tikzpicture}
	}
	\subfloat[Roomset-noDC]{
		\begin{tikzpicture}
			\begin{axis}[
				width = 0.25\textwidth,
				height = 0.23\textwidth,
				axis y line = left,
				axis x line = bottom,
				xtick={1,2,3,4,5,6},
				xticklabels={bass,piano,voice,drums,drips,nature},
				x tick label style={rotate=90},
				ytick={},
				xlabel={},
				ylabel={$S$/sone},
				xmin = 0.5, xmax = 6.5,
				ymin = 0, ymax = 15,
				/pgf/number format/.cd
			]
			\addplot[blue-base, mark=*, mark size=0.5pt, mark layer=axis background, only marks] table [x=input,y=loudness_0, unbounded coords=discard, restrict expr to domain={((\thisrow{scene} - 4) * 2 + (\thisrow{flag} - 1)) * 10 + \thisrow{tester}}{1:8}, unbounded coords=discard, col sep=comma] {img/actual_sound_loudness_ag_hi.csv};
			\addplot[gray-base-dark, mark=*, mark size=1pt, mark layer=axis background, only marks] table [x=input,y=loudness_0, unbounded coords=discard, restrict expr to domain={((\thisrow{scene} - 4) * 2 + (\thisrow{flag} - 1)) * 10 + \thisrow{tester}}{9:10}, unbounded coords=discard, col sep=comma] {img/actual_sound_loudness_ag_hi.csv};
			\addplot[red-base, mark=*, mark size=0.5pt] table [x=input,y=avg, unbounded coords=discard, restrict expr to domain={(\thisrow{scene} - 4)}{0:0}, unbounded coords=discard, col sep=comma] {img/actual_sound_loudness_ag_hi_avg.csv};
			\end{axis}
		\end{tikzpicture}
	}
	\caption{Error loudness at measured audibility thresholds for the 28-55 age group with hearing-impaired subjects. The test conditions are identical to those of \autoref{FIG:THRESHOLD-BY-CASES}. The measured threshold loudness values are marked with blue dots for subjects with normal hearing, and black dots for hearing impaired subjects. The average thresholds for different input audios are connected with red lines.}
	\label{FIG:THRESHOLD-BY-CASES2}
\end{figure*}
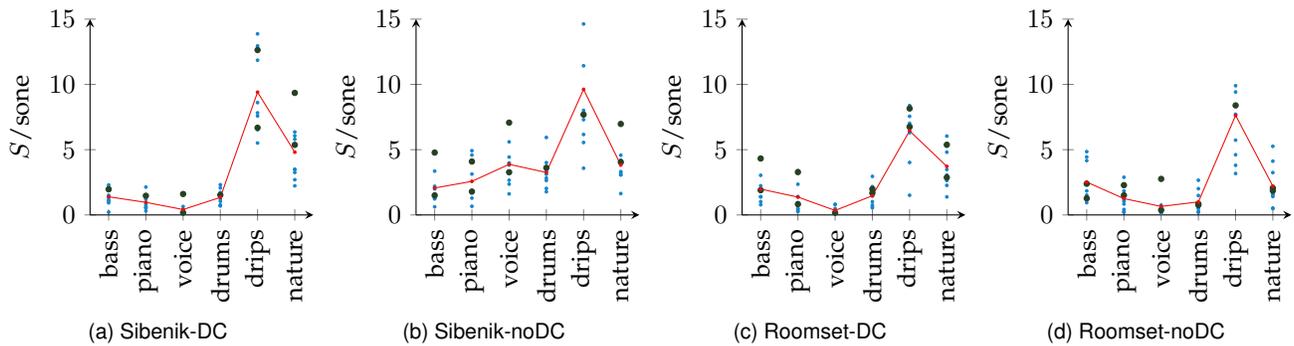

Our validation tests on real outputs consist of 2 subgroups:
\begin{enumerate}
	\item Validation on 30 subjects from the age group 18-27.
	\item Validation on 10 subjects from the age group 28-55.
\end{enumerate}
The test audio for both subgroups is generated with the BDPT simulator. To detect and remove the outliers in our measured data, we use the generalized ESD test \cite{Rosner1983PercentagePF}, which is effective for small-sample, normally distributed datasets with multiple outliers. We set the maximum number of outliers to 5 for the first subgroup and 3 for the second one. The results of our measurement are presented in \autoref{FIG:THRESHOLD-BY-CASES} and \autoref{FIG:THRESHOLD-BY-CASES2}.

\begin{figure}
	\centering
	\begin{tikzpicture}
		\begin{axis}[
			width = 0.4\textwidth,
			height = 0.24\textwidth,
			axis y line = left,
			axis x line = bottom,
			xlabel={IR energy SNR / dB},
			ylabel={Error Loudness / sone},
			xmin = -10, xmax = 30,
			ymin = 0, ymax = 3,
			legend style = {draw=none, fill=none},
			/pgf/number format/.cd
		]
		\addplot[blue-base] table [x=SNR,y=bass, col sep=comma] {img/loudness_SNR_relation.csv};
		\addlegendentry{bass};
		\addplot[red-base] table [x=SNR,y=piano, col sep=comma] {img/loudness_SNR_relation.csv};
		\addlegendentry{piano};
		\addplot[green-base] table [x=SNR,y=voice, col sep=comma] {img/loudness_SNR_relation.csv};
		\addlegendentry{voice};
		\addplot[blue-base,mark=*, mark size=1pt, only marks] table [x=SNR,y=loudness, col sep=comma] {img/threshold_bass.csv};
		\addplot[red-base,mark=*, mark size=1pt, only marks] table [x=SNR,y=loudness, col sep=comma] {img/threshold_piano.csv};
		\addplot[green-base,mark=*, mark size=1pt, only marks] table [x=SNR,y=loudness, col sep=comma] {img/threshold_voice.csv};
		\end{axis}
	\end{tikzpicture}
	\caption{Relationship between IR energy SNR and noise loudness of three melodic audios under the ``Sibenik-DC'' scene configuration. The IRs are generated with the BDPT simulator. The measured error audibility thresholds of 30 subjects (age group 18-27) are marked on the loudness curves.}
	\label{FIG:LOUDNESS-SNR-RELATION}
\end{figure}
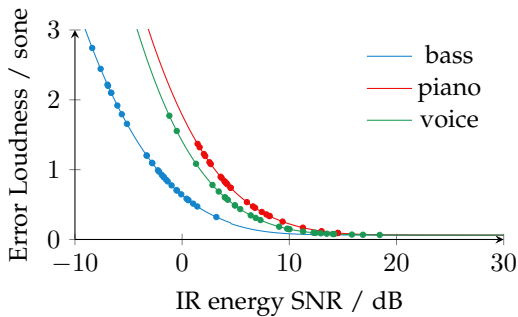

There may be concerns about the robustness of our criterion to different inputs. Ideally, we expect the measured error audibility thresholds to be ``nearly equally loud'' in all cases. \autoref{FIG:THRESHOLD-BY-CASES} show that the average audibility thresholds are largely stable for different acoustic environments and subject age groups. We can also see from \autoref{FIG:THRESHOLD-BY-CASES2} that for hearing-impaired subjects, the measured thresholds are not particularly different from those of other subjects, and our criterion remains valid.

\autoref{FIG:LOUDNESS-SNR-RELATION} shows that our criterion can help in the choice of IR SNR for the actual rendering process. The figure shows the loudness-SNR relationship curves under a certain scene configuration, and the measured noise audibility thresholds from subgroup 1 are marked on the curves. We can see that the error audibility thresholds cluster around the ``knee'' part of the relationship curves, where the loudness goes up quickly when the SNR decreases and stay invariant when the SNR increases. Our criterion can explain the difference of error significance between different input audios very well. The difference between the average loudness thresholds of the three input audios is less than 0.43 sone. However, the average IR SNR required for the error to be imperceptible is -3.29 dB when the input is ``bass'', and 4.47 dB when the input is ``piano'', which is much higher. While it is known that human hearing is less sensitive to the difference of low-frequency sounds, our criterion can show this difference in a quantitative manner.

We can also see some other interesting phenomena in the results. The noise of all the ``melodic'' input audios is almost equally perceptible. The average noise loudness threshold for the melodic inputs is 1.308, and 83\% of the measured thresholds fall between 0-2 sone. On the other hand, the noise of input ``drips'' and ``nature'' is much more difficult to perceive than other inputs. While the input ``drums'' is not melodic, its noise perceptibility is closer to melodic audios than noisy audios. These phenomena probably indicate that semantic information is an important cue for error perception: the other two noisy audios are very close to ``random noise'' while the semantic richness of the ``drums'' audio is similar to melodic audios.

\subsubsection{Generalizability for Different Simulation Algorithms}

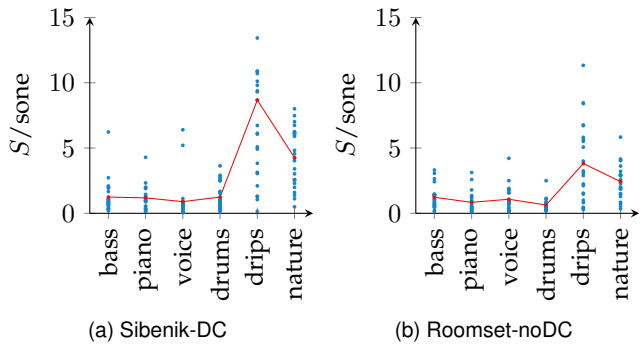
\begin{figure}[htbp]
	\centering
	\subfloat[Sibenik-DC]{
		\begin{tikzpicture}
			\begin{axis}[
				width = 0.25\textwidth,
				height = 0.23\textwidth,
				axis y line = left,
				axis x line = bottom,
				xtick={1,2,3,4,5,6},
				xticklabels={bass,piano,voice,drums,drips,nature},
				x tick label style={rotate=90},
				ytick={},
				xlabel={},
				ylabel={$S$/sone},
				xmin = 0.5, xmax = 6.5,
				ymin = 0, ymax = 15,
				/pgf/number format/.cd
			]
			\addplot[blue-base, mark=*, mark size=0.5pt, mark layer=axis background, only marks] table [x=input,y=loudness_0, unbounded coords=discard, restrict expr to domain={((\thisrow{scene} - 1) * 2 + (\thisrow{flag} - 1)) * 25 + \thisrow{tester}}{1:25}, unbounded coords=discard, col sep=comma] {img/actual_sound_loudness_dis.csv};
			\addplot[red-base, mark=*, mark size=0.5pt] table [x=input,y=avg, unbounded coords=discard, restrict expr to domain={(\thisrow{scene} - 1)}{0:0}, unbounded coords=discard, col sep=comma] {img/actual_sound_loudness_dis_avg.csv};
			\end{axis}
		\end{tikzpicture}
	}
	\subfloat[Roomset-noDC]{
		\begin{tikzpicture}
			\begin{axis}[
				width = 0.25\textwidth,
				height = 0.23\textwidth,
				axis y line = left,
				axis x line = bottom,
				xtick={1,2,3,4,5,6},
				xticklabels={bass,piano,voice,drums,drips,nature},
				x tick label style={rotate=90},
				ytick={},
				xlabel={},
				ylabel={$S$/sone},
				xmin = 0.5, xmax = 6.5,
				ymin = 0, ymax = 15,
				/pgf/number format/.cd
			]
			\addplot[blue-base, mark=*, mark size=0.5pt, mark layer=axis background, only marks] table [x=input,y=loudness_0, unbounded coords=discard, restrict expr to domain={((\thisrow{scene} - 2) * 2 + (\thisrow{flag} - 1)) * 25 + \thisrow{tester}}{1:25}, unbounded coords=discard, col sep=comma] {img/actual_sound_loudness_dis.csv};
			\addplot[red-base, mark=*, mark size=0.5pt] table [x=input,y=avg, unbounded coords=discard, restrict expr to domain={(\thisrow{scene} - 2)}{0:0}, unbounded coords=discard, col sep=comma] {img/actual_sound_loudness_dis_avg.csv};
			\end{axis}
		\end{tikzpicture}
	}
	\caption{Error loudness at measured audibility thresholds for 25 subjects aged 18-55. The audio is generated with the HIS simulator. The measured threshold loudness values are marked with blue dots, and the average thresholds for different input audios are connected with red lines.}
	\label{FIG:THRESHOLD-BY-CASES3}
\end{figure}

To validate the generalizability of our criterion, we performed an additional test on 25 subjects (20 from the age group 18-27, 5 from the age group 28-55) with audios generated by the HIS simulator. The result is presented in \autoref{FIG:THRESHOLD-BY-CASES3}. We can see from the figures that the HIS result is consistent with the BDPT result in \autoref{FIG:THRESHOLD-BY-CASES} and \autoref{FIG:THRESHOLD-BY-CASES2}.

\section{Conclusion and Future Work}

In this paper, we demonstrated that an algorithm that calculates the auditory effects of sound propagation simulation error can be built on a solid theoretical basis. Experimental results show that our criterion is valid for various sounds and people with different hearing abilities, and the result matches those of previous research. Our spectrum analysis makes it possible to use complex psychoacoustic models in sound propagation error control and opens the possibility for further application of psychoacoustic tools in the field of sound propagation.

Our error analysis for path tracing relies only on the assumption of sample independence. Thus it can be applied to other path traced methods. Our analysis is also helpful in the optimization of sound propagation simulators. For example, \autoref{eq:fullerrorspectrum} shows us the connection between the shape of the interpolation window and the error spectrum. Many audio renderers use linear interpolation between frames, which is equivalent to using a triangular window. Our analysis shows that a better window function, like the Hanning window, could improve the final quality of the output audio.

There is still a lot of work to do related to the current criterion. We will list a few possible directions here.

We have mentioned that Zwicker's loudness algorithm has another version for ``time-varying'' sounds, the spectra of which vary significantly during playback. One can see from \autoref{FIG:THRESHOLD-BY-CASES} that the noise loudness for input audio ``drips'' seems to be very different from other audios. The audio ``drips'' contains intermittent water dripping sounds and is not as ``stationary'' as other inputs. This probably indicates that we need to modify our algorithm when dealing with time-varying input audios.

The discussion in \autoref{section:real-outputs} shows that audio semantics is possibly another important factor influencing error perception. To analyze the influence of semantics, we need to model the deep-level mechanisms of human auditory cognition, which is a good topic for future research.

While our ``soft threshold'' has taken the variance of human hearing abilities into consideration, the variance model is probably oversimplified. We chose the logistic distribution in \autoref{SEC:SOFT-THRESHOLD} for its mathematical simplicity. Ideally, for a better description of $k_l$, we need a nonsymmetric distribution defined on $[0,+\infty)$ that is relatively uniform on the logarithmic scale. Unfortunately, distributions satisfying such conditions usually have very complex CDFs with no analytical integral expression. This problem could be solved with numerical approximation, from which we may derive a soft threshold that better explains human hearing. 

Finally, while we expect our criterion to be able to improve the efficiency of existing sound simulators, especially the real-time ones, it is not as simple as it seems. Our criterion uses signal statistics and operations on the spectral domain. The statistics require knowledge of multiple IR frames, and the Fourier transform can be time-consuming. To integrate our criterion into real applications, we need an efficient implementation that works well with real-time simulators, which we will try to develop in the future.

\bibliographystyle{IEEEtran}
\bibliography{cites.bib}


\begin{IEEEbiography}[{\includegraphics[width=1in,height=1.25in,clip,keepaspectratio]{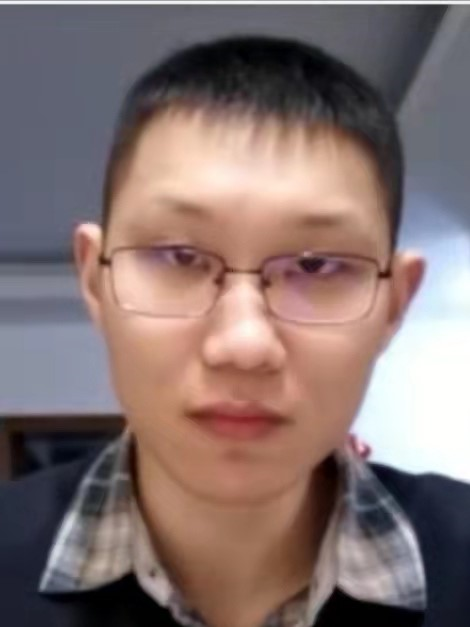}}]{Chunxiao Cao}
	received the bachelor’s degree in mathematics from Zhejiang University in 2014. He is currently working toward the PhD degree in the Graphics and Parallel Systems Lab of Zhejiang University. His research interests include sound propagation simulation in combination with computer graphics techniques.
\end{IEEEbiography}

\begin{IEEEbiography}[{\includegraphics[width=1in,height=1.25in,clip,keepaspectratio]{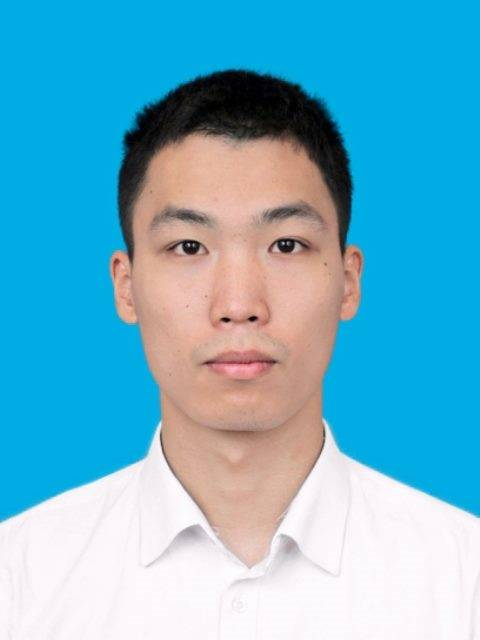}}]{Zili An}
	received the bachelor’s degree in information and computer science from Zhejiang University in 2016. He is currently working toward the master degree in the Graphics and Parallel Systems Lab of Zhejiang University. His research interests include sound rendering.
	\end{IEEEbiography}

\begin{IEEEbiography}[{\includegraphics[width=1in,height=1.25in,clip,keepaspectratio]{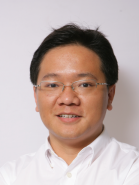}}]{Zhong Ren}
	received the bachelor’s degree in information engineering, the master’s degree in mechanical engineering, and the PhD degree in computer science in 2007 all from Zhejiang University. He is currently an associate professor at Zhejiang University, and a member of the State Key Lab of CAD\&CG. Before returning to Zhejiang University in 2010, he was at Microsoft Research Asia for three years, where he was an associate researcher in the Internet Graphics Group. His research interests include real-time rendering of soft shadows and participating media, spherical harmonics for real-time rendering, and GPU-based photorealistic rendering.
\end{IEEEbiography}

\begin{IEEEbiography}[{\includegraphics[width=1in,height=1.25in,clip,keepaspectratio]{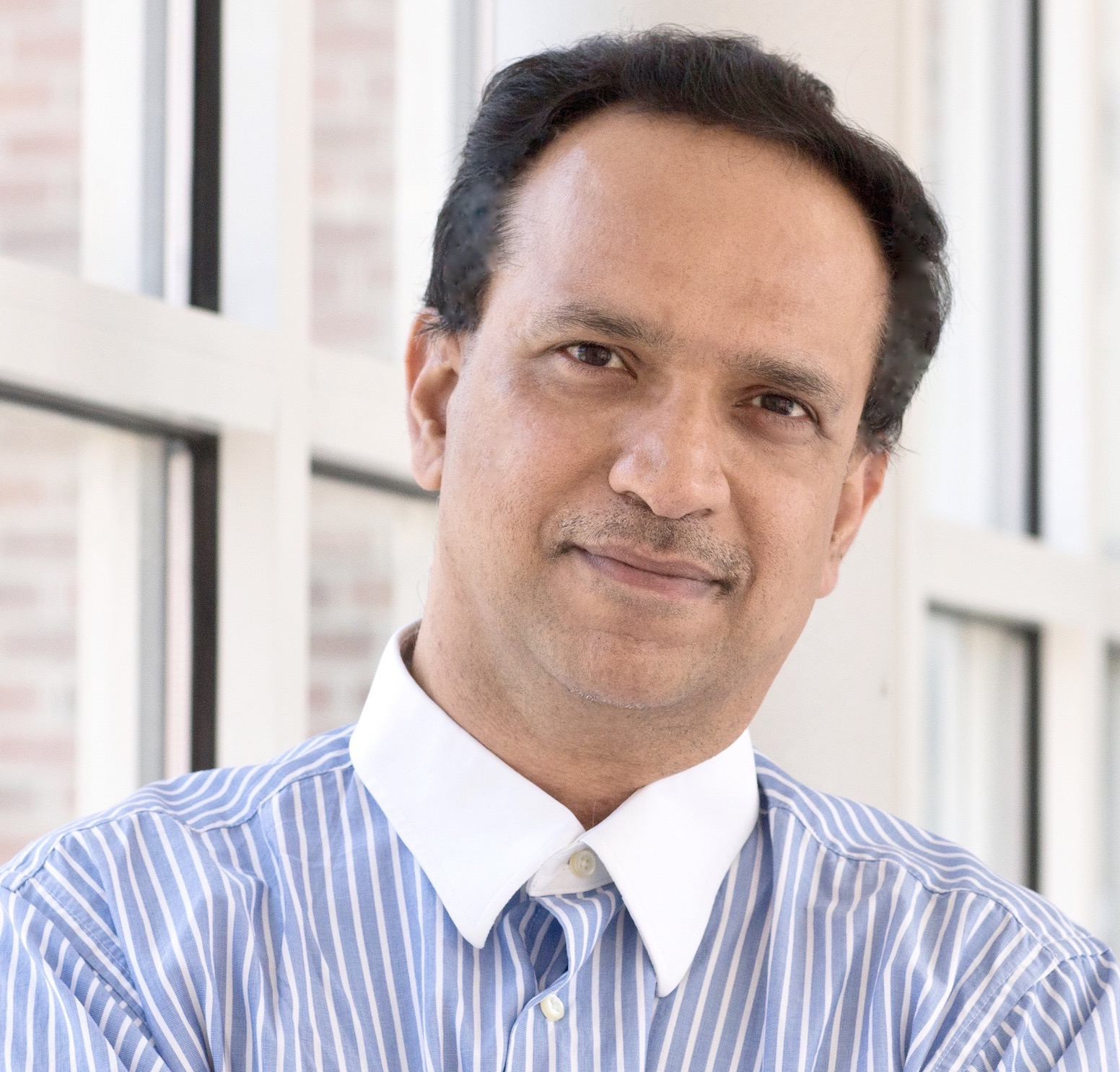}}]{Dinesh Manocha}
	is a Distinguished University Professor of the University of Maryland, where he is the Paul Chrisman Iribe Professor of Computer Science and Professor of Electrical and Computer Engineering. Manocha’s research focuses on AI and robotics, computer graphics, augmented/virtual reality, and  scientific computing.  He has published more than 600 papers in these areas and won 17 best paper awards at leading conferences. His group has  developed a number of software technologies that are licensed to more than 60 commercial vendors. A  Fellow of AAAI, AAAS, ACM, IEEE and Sloan Foundation,  Manocha is a member of ACM SIGGRAPH Academy Class, and Bézier Award recipient from Solid Modeling Association. He received the Distinguished Alumni Award from IIT Delhi and the Distinguished Career in Computer Science Award from Washington Academy of Sciences. He was also the co-founder of Impulsonic, a developer of physics-based audio simulation technologies, which was acquired by Valve Inc in November 2016.
\end{IEEEbiography}

\begin{IEEEbiography}[{\includegraphics[width=1in,height=1.25in,clip,keepaspectratio]{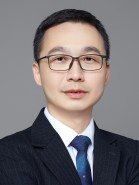}}]{Kun Zhou}
	received the BS and PhD degrees in computer science from Zhejiang University, in 1997 and 2002, respectively. He is currently a Cheung Kong distinguished professor in the Computer Science Department of Zhejiang University, and a member of the State Key Lab of CAD\&CG. Before joining Zhejiang University, he was a lead researcher of the Graphics Group at Microsoft Research Asia. His research interests include shape modelling/editing, texture mapping/synthesis, animation, rendering, and GPU parallel computing. He is a fellow of IEEE.
\end{IEEEbiography}

\vfill


\end{document}